\documentclass[floatfix, twocolumn, tighten]{aastex63}

\newcommand{\change}[1]{#1}
\newcommand{\changeref}[1]{#1}
\newcommand{\colorVIACS}{\ensuremath{m_{V_{F606W}} - m_{\rm I_{F814W}}}}
\newcommand{\colorVI}{\ensuremath{m_{V} - m_{\rm I}}}
\newcommand{\bigcell}[2]{\begin{tabular}{@{}#1@{}}#2\end{tabular}}
\usepackage{multirow}
\usepackage{amsmath}
\received{}
\revised{}
\accepted{}

\submitjournal{ApJ}
\shorttitle{}
\begin{document}

\title{Hyper Suprime-Cam Low Surface Brightness Galaxies II: A Hubble Space Telescope Study of the Globular Cluster Systems of Ultra-Diffuse Galaxies in Groups\footnote{Based on observations made with the NASA/ESA Hubble Space Telescope, obtained at the Space Telescope Science Institute, which is operated by the  Association of Universities for Research in Astronomy, Inc., under NASA contract NAS 5-26555. These observations are associated with programs GO-15277}}

\author[0000-0001-8426-5732]{Jean J. Somalwar}
\affil{Department of Astrophysical Sciences, 4 Ivy Lane, Princeton University, Princeton, NJ 08544, USA}

\author{Jenny E. Greene}
\affil{Department of Astrophysical Sciences, 4 Ivy Lane, Princeton University, Princeton, NJ 08544, USA}

\author[0000-0003-4970-2874]{Johnny P. Greco}
\altaffiliation{NSF Astronomy \& Astrophysics Postdoctoral Fellow}
\affiliation{Center for Cosmology and AstroParticle Physics (CCAPP), The Ohio State University, Columbus, OH 43210, USA}

\author{Song Huang}
\affil{Department of Astrophysical Sciences, 4 Ivy Lane, Princeton University, Princeton, NJ 08544, USA}

\author{Rachael L. Beaton}
\affil{Department of Astrophysical Sciences, 4 Ivy Lane, Princeton University, Princeton, NJ 08544, USA}

\author{Andy D. Goulding}
\affil{Department of Astrophysical Sciences, 4 Ivy Lane, Princeton University, Princeton, NJ 08544, USA}

\author[0000-0002-0041-4356]{Lachlan Lancaster}
\affil{Department of Astrophysical Sciences, 4 Ivy Lane, Princeton University, Princeton, NJ 08544, USA}

\correspondingauthor{Jean J. Somalwar}
\email{jsomalwar@gmail.com}

\begin{abstract}
We increase the sample of ultra diffuse galaxies (UDGs) in lower density environments with characterized globular cluster (GC) populations using new {\it Hubble Space Telescope} observations of nine UDGs in group environments. While the bulk of our UDGs have GC abundances consistent with normal dwarf galaxies, two of these UDGs have excess GC populations. These two UDGs both have GC luminosity functions consistent with higher surface brightness galaxies and cluster UDGs. We then combine our nine objects with previous studies to create a catalog of UDGs with analyzed GC populations that spans a uniquely diverse range of environments. We use this catalog to examine broader trends in the GC populations of low stellar mass galaxies. The highest GC abundances are found in cluster UDGs, but whether cluster UDGs are actually more extreme requires study of many more UDGs in groups. We find a possible positive correlation between GC abundance and stellar mass, and between GC abundance and galaxy size at fixed stellar mass. However, we see no significant stellar-mass galaxy-size relation, over our limited stellar mass range. We consider possible origins of the correlation between GC abundance and galaxy size, including the possibility that these two galaxy properties are both dependent on the galaxy dark matter halo, or that they are related through baryonic processes like internal feedback.
\end{abstract}

\section{Introduction} \label{sec:intro}
Low surface brightness galaxies (LSBGs) are a powerful probe of the coupling between baryons and the dark matter halos they inhabit. Ultra-diffuse galaxies (UDGs), which are characterized by central surface brightnesses $\mu_{0,g} > 24$ mag arcsec$^{-2}$ and sizes $r_{\rm eff} > 1.5$ kpc, are especially sensitive to the astrophysics of star formation and feedback. \changeref{UDGs have been known to exist since the 1980s \citep[e.g.][]{Sandage1984, Impey1988, Dalcanton1997,Conselice2003GalaxyPopulations}, but it has only recently become possible to build large samples of such low surface brightness objects across galaxy environments using deep, wide-field imaging surveys. The discovery of a large population of UDGs in the Coma Cluster by the low-surface-brightness-optimized Dragonfly Array \citep{Abraham2014UltraLowArray, VanDokkum2015} has reignited the search for such objects, leading to the discovery of thousands of UDGs in Coma and other clusters \citep[][]{Koda2015APPROXIMATELYCLUSTER, Yagi2016,VanDerBurg2016TheClusters, Lee2020TheFields, Pina2019ThetextttKIWICS}, as well as in lower density environments \cite[e.g.][]{Leisman2017AlmostGalaxies, Greco2018IlluminatingSurvey, Roman2017, VanDerBurg2017TheHaloes}.}

Despite this monumental increase in sample size, many puzzles surround UDGs; in particular, their formation mechanism is unclear. We can separate the proposed UDG formation models by their predicted halo masses. At the typical UDG stellar mass (${\sim}10^8\,{M}_{\odot}$), we expect halo masses $M_{\rm halo} \sim 10^{10-11}\,{M}_{\odot}$. If UDGs occupy this $M_{\rm halo}$ range, they may be outliers in their surface brightness due to high angular momentum halos, strong stellar outflows, or tidal stripping \citep[e.g.][]{Amorisco2016, Carleton2019TheHeating, Liao2019Ultra-diffuseSimulations, DiCintio2017, Jiang2019FormationGroups, El-Badry2016BreathingGalaxies}. On the other hand, if UDGs have large halo masses $M_{\rm halo}\sim10^{11-12}\,{M}_{\odot}$, they may be produced by, e.g., the ``failed-$L_*$'' mechanism, which forms UDGs as normal $L_*$ galaxies which suffer early quenching \citep[][]{VanDokkum2015, Yozin2015}.

This large variation in possible halo masses for UDGs reflects the complex nature of the galaxy-halo connection in the dwarf regime, which we can quantify with the stellar-to-halo mass relation (SHMR) \citep{Wechsler2018TheHalos}. 
The SHMR is a correlation between $M_*$ and $M_{\rm halo}$. It is often modelled as a broken power law with a break near $M_* \sim 10^{12}\,{M}_{\odot}$ and lognormal scatter. At low $M_*$ ($\lesssim 10^{8-9}\,{M}_{\odot}$), observations are limited and the slope and scatter of the SHMR become very degenerate, so the level of scatter is heavily debated \citep[e.g.][]{Garrison-Kimmel2016OrganizedGalaxies, Cao2019ConstrainingMasses}. 

Therefore, it is important to constrain the halo masses of UDGs to better understand their formation, and more broadly how they affect the SHMR at low stellar masses. There are two main avenues to better understanding UDG halo masses. First, cosmological simulations contain UDG-like galaxies that are roughly consistent with the standard SHMR, but not objects in overly massive halos \citep[][]{Carleton2019TheHeating, Liang:2016, Jiang2019FormationGroups, DiCintio2017}. Those simulations which contain $M_{\rm halo} \lesssim 10^{11}\,{M}_{\odot}$ UDGs disagree about the precise mechanism driving the large galaxy effective radii. For example, simulations disagree about the importance of halo spin in forming UDGs \citep[][]{Carleton2019TheHeating, Liao2019Ultra-diffuseSimulations, DiCintio2017, Tremmel2019TheSimulation}.

\begin{deluxetable*}{ccccccccc}
\tablenum{1}
\tablecaption{Summary of ultra-diffuse galaxy properties  \label{tab:gals}}
\tablewidth{0pt}
\tablehead{
\colhead{Galaxy} & \colhead{RA [J2000]} & \colhead{Decl. [J2000]} & \colhead{$m_{V}$ [mag]} & \colhead{$\mu_0(g)$ [mag arcsec$^{-2}$]} & \colhead{$m_{\rm g}$ [mag]} & \colhead{$r_{\rm eff}$ [arcsec]} & \colhead{$r_{\rm eff}$ [kpc]} & \colhead{$\log\,M_*/{M}_{\odot}$}}
\startdata
UDG-1A & $09{:}18{:}45.32$ & $+00{:}24{:}01.40$ & $19.5 \pm 0.4$ & $24.1 \pm 0.2$ & $19.9 \pm 0.2$ & $6.2 \pm 0.8$ & $2.1 \pm 0.3$ & $7.9 \pm 0.3$ \\ 
UDG-3A & $09{:}19{:}55.56$ & $+01{:}07{:}23.77$ & $20.0 \pm 0.4$ & $24.2 \pm 0.2$ & $20.3 \pm 0.2$ & $4.6 \pm 0.8$ & $1.6 \pm 0.3$ & $7.7 \pm 0.3$ \\
UDG-4A & $09{:}18{:}55.41$ & $+01{:}45{:}05.69$ & $19.7 \pm 0.4$ & $24.9 \pm 0.2$ & $20.0 \pm 0.2$ & $9.5 \pm 0.8$ & $3.3 \pm 0.3$ & $7.8 \pm 0.3$ \\
UDG-5A & $09{:}19{:}59.20$ & $+00{:}48{:}52.63$ & $20.1 \pm 0.4$ & $25.6 \pm 0.2$ & $20.2 \pm 0.2$ & $8.2 \pm 0.8$ & $2.9 \pm 0.3$ & $7.6 \pm 0.3$ \\
\hline
UDG-1B & $12{:}04{:}46.26$ & $+01{:}17{:}54.20$ & $19.6 \pm 0.4$ & $23.8 \pm 0.2$ & $19.6 \pm 0.2$ & $5.2 \pm 0.8$ & $2.2 \pm 0.3$ & $8.0 \pm 0.3$ \\ 
UDG-2B & $12{:}04{:}07.86$ & $+01{:}19{:}17.62$ & $20.1 \pm 0.4$ & $24.1 \pm 0.2$ & $20.0 \pm 0.2$ & $5.5 \pm 0.8$ & $2.3 \pm 0.3$ & $7.8 \pm 0.3$ \\ 
UDG-3B & $12{:}04{:}33.60$ & $+01{:}24{:}57.31$ & $20.1 \pm 0.4$ & $24.2 \pm 0.2$ & $20.0 \pm 0.2$ & $4.7 \pm 0.8$ & $2.0 \pm 0.3$ & $7.8 \pm 0.3$ \\
UDG-4B & $12{:}02{:}37.06$ & $+01{:}30{:}27.00$ & $20.0 \pm 0.4$ & $24.7 \pm 0.2$ & $19.9 \pm 0.2$ & $7.6 \pm 0.8$ & $3.2 \pm 0.3$ & $7.8 \pm 0.3$ \\
UDG-5B & $12{:}06{:}30.00$ & $+01{:}33{:}22.75$ & $21.4 \pm 0.4$ & $26.0 \pm 0.2$ & $21.1 \pm 0.2$ & $5.4 \pm 0.8$ & $2.2 \pm 0.3$ & $7.3 \pm 0.3$ \\
\enddata
\tablecomments{Galaxy properties determined from the \cite{Greco2018IlluminatingSurvey} catalog. We assume a distance of ${\sim}75$ (90) Mpc for UDGs in group A (B) \citep{Yang2007}. The $V$-band magnitude, $M_{V}$, is calculated as $M_{V} = M_g - 0.59(M_g - M_r)-0.01$ \citep[][]{Jester2005}. 
The stellar masses are calculated assuming a solar mass-to-light ratio. We assume that the stellar mass error is $0.3$ dex.}
\end{deluxetable*}

Second, UDG halo masses can be constrained observationally using measurements of weak lensing \citep[][]{Sifon2018ALensing}, dynamics \citep[][]{Beasley2016a, VanDokkum2017,VanDokkum2018AMatter, VanDokkum2018, Danieli2019AData}, and X-ray emission \citep[][]{Kovacs2019ConstrainingGalaxies, Kovacs2020Ultra-diffuseFraction}. Such work has suggested that UDGs have halo masses covering the entire range from $M_{\rm halo} \sim 10^{8}\,{M}_{\odot}$ to $M_{\rm halo} \sim 10^{12}\,{M}_{\odot}$, or from dwarf masses to $L_*$ or larger. \citep[e.g.][]{VanDokkum2017, VanDokkum2018AMatter, VanDokkum2018, Danieli2019AData}. However, due to the technical challenges involved with observing low surface brightness objects, these methods of constraining halo masses have only yielded individual halo masses for a small number of targets.

Alternatively, one can potentially measure halo masses with globular cluster (GC) abundance, which we will refer to as $N_{\rm GC}$. A correlation between $N_{\rm GC}$ and $M_{\rm halo}$ is suggested by the U-shaped distribution of $N_{\rm GC}$ as a function of galaxy stellar mass \citep[e.g.][]{Harris1981GlobularEllipticals, Peng2008, Blakeslee1997}. Such a correlation has been observed in high surface brightness galaxies \citep[][]{Harris2013, Harris2017}, \changeref{ and comparisons of dynamical UDG masses to GC counts have suggested that the same correlation is valid in the UDG regime \citep[][]{Beasley2016a, Toloba2018}}. GCs are gravitationally bound clumps of stars which form early and have typical masses $M_*\sim 10^{5}\,{M}_{\odot}$ and mass-to-light ratios $(M_*/L_{V})/{ (M_*/L_V)}_\odot \sim 2$ \citep[][]{Ebrahimi2020NewClustersc}. They have a typical size ${\sim}10$ pc, corresponding to ${\sim}0\farcs02$ at a distance ${\sim}100$ Mpc \citep[][]{Kruijssen2014GlobularEvolution}. The theoretical basis for the connection between GC abundance and dark matter halo mass is hotly debated. Simulations can reproduce the correlation by invoking tidal disruption in the dense disks where GCs form \citep[e.g.][]{Kruijssen2015GlobularGalaxies}, but it has also been reproduced at $M_{\rm halo}\gtrsim 10^{11.5}\,{M}_\odot$ using hierarchical merging, with no dependence on the GC formation history \citep{El-Badry2019ThePopulations}.

Although some UDGs have small GC populations consistent with dwarf galaxies \citep[e.g.][]{Amorisco2016a, Forbes2018}, a number of UDGs have very rich GC populations \citep[e.g.][]{Peng2016, VanDokkum2017}, which could suggest that the UDGs are produced by a mechanism which also predicts high halo masses. However, the objects with large GC populations are all located in clusters, raising the question of whether GC-rich UDGs are unique to high density environments. The LSBG catalog from \cite{Greco2018IlluminatingSurvey} provides a unique opportunity to explore this possibility with its large environment- and color-blind sample of LSBGs. Thus, in this paper we will use {\it Hubble Space Telescope} (\emph{HST}) observations of nine \cite{Greco2018IlluminatingSurvey} UDGs in group environments to constrain the GC populations of UDGs in lower density environments \citep{Greco2017WeighingGap}. 

In Section~\ref{sec:obs}, we describe our observations and data reduction procedure. In Section~\ref{sec:GCdet}, we describe our procedure to identify GCs associated with our UDGs. In Section~\ref{sec:results}, we describe the GC populations of our UDGs. Then, we place our work in a broader context by comparing our observations with literature results in Section~\ref{sec:trends}. We will analyze the relationship between GC abundance, galaxy stellar mass, and galaxy size in Section~\ref{sec:covar}. Finally, we will summarize and conclude in Section~\ref{sec:discussion}.

We adopt a standard flat $\Lambda$CDM model with H$_0 = 70$ km s$^{-1}$ Mpc$^{-1}$ and $\Omega_{\rm m} = 0.3$. All magnitudes are reported in the AB system. 

\section{Data} \label{sec:obs}

\subsection{Sample Selection}
\begin{figure*}
\gridline{\fig{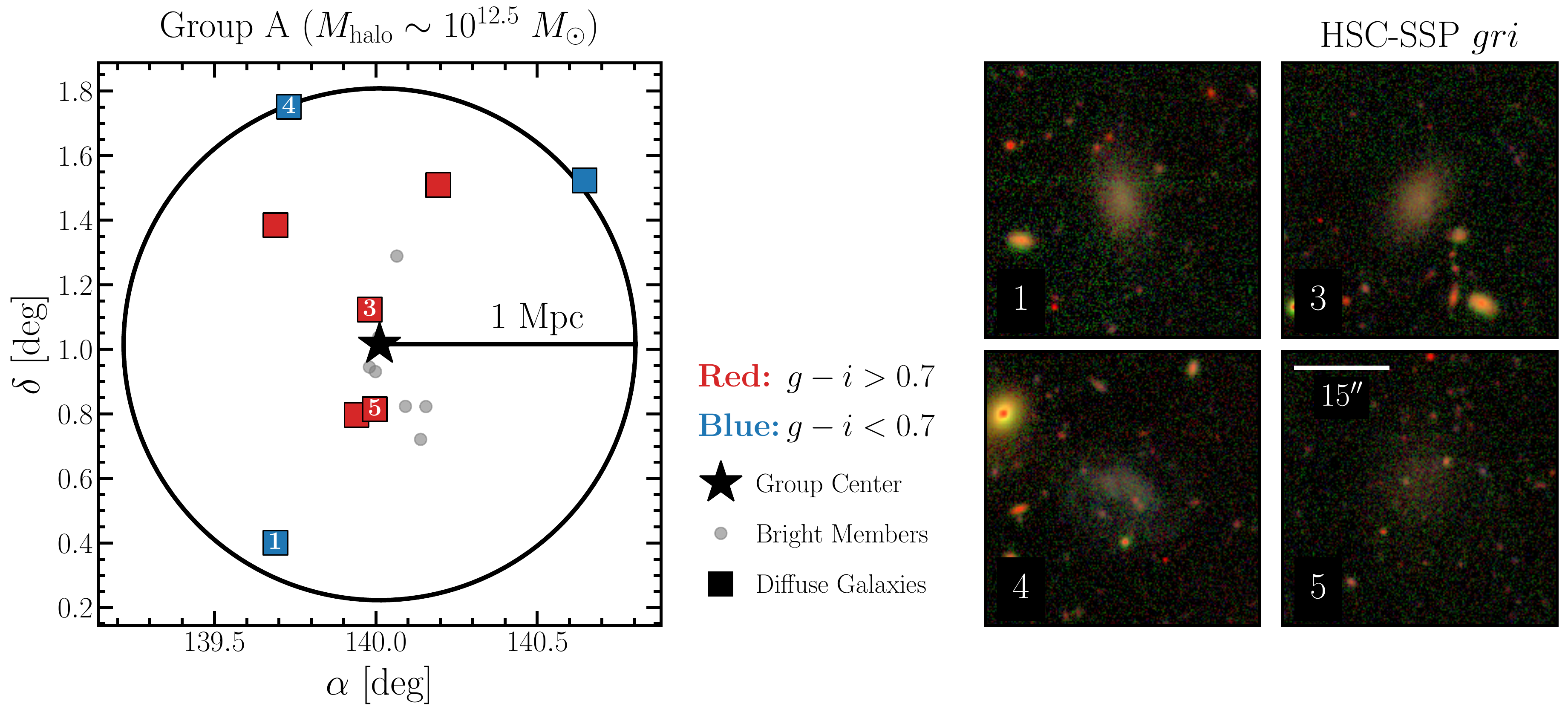}{\textwidth}{}}
\gridline{\fig{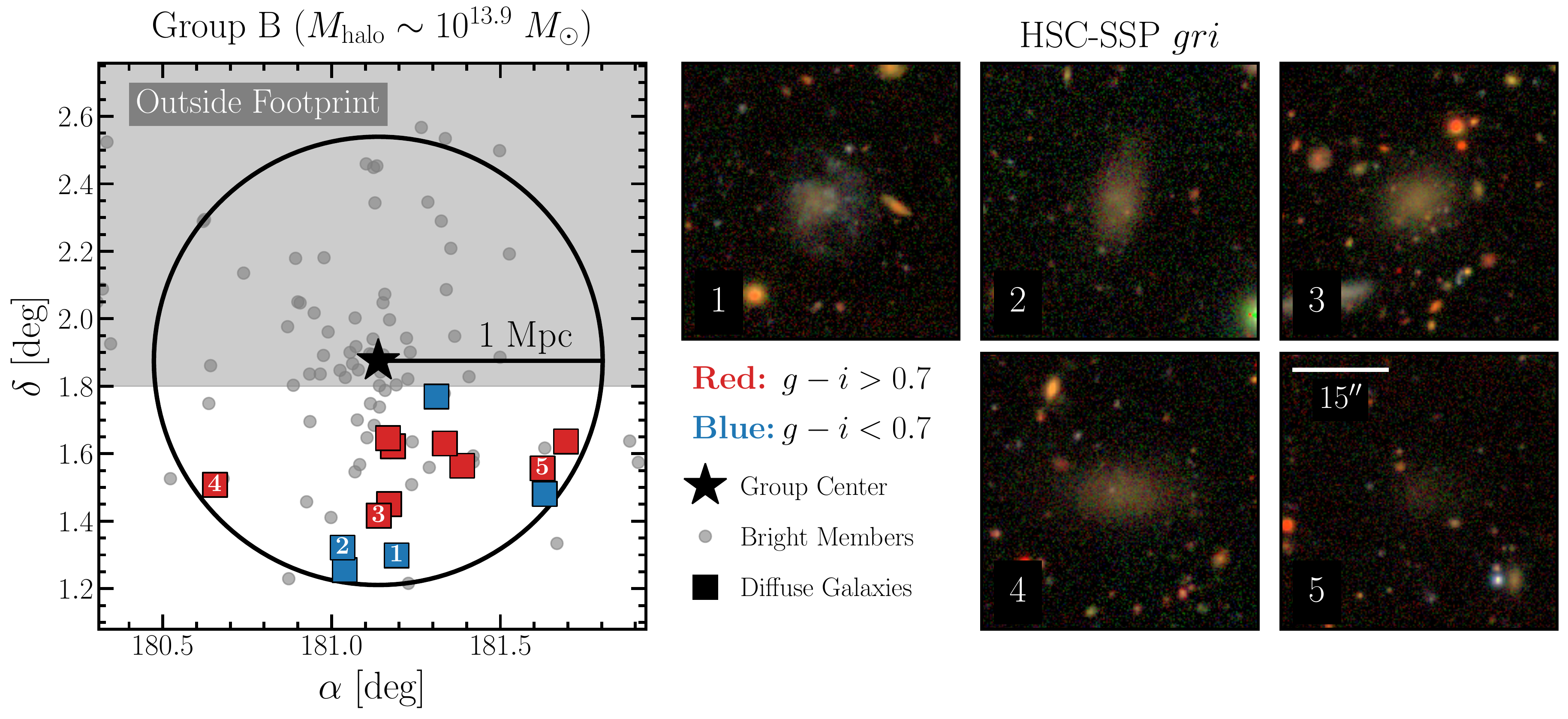}{\textwidth}{}}
\caption{ Summary of the UDG sample selection. The UDGs are contained in galaxy groups A ($z = 0.017$, host halo mass $M_{\rm host} = 10^{12.5}\,{M}_{\odot}$) and B ($z = 0.0206$, $M_{\rm host} = 10^{13.9}\,{M}_{\odot}$). The {\it left} panels show the placement of galaxies within group A ({\it top}) and group B ({\it bottom}). UDGs are shown as squares with color indicating the galaxy color, and galaxies from the SDSS catalogues are shown as gray dots. The UDGs analyzed in this work are numbered. The {\it right} panels show color Hyper Suprime-Cam images of the nine UDGs analyzed in this work. \label{figure:galgrid_color}}
\end{figure*}

We selected a sample of UDGs in two groups from the \cite{Greco2018IlluminatingSurvey} catalog. \cite{Greco2018IlluminatingSurvey} designed a custom search of Hyper Suprime-Cam (HSC) data and identified ${\sim}800$ LSBGs over ${\sim}200$ deg$^2$ \citep{Bosch2018ThePipeline, Aihara2018FirstProgram}. Within this parent sample, we selected UDGs projected near two galaxy groups of known distance. These groups were identified in the SDSS galaxy groups survey using a friends-of-friends halo finder, and were assigned halo masses based on total $r$-band luminosity \citep[][]{Yang2007}. Of the UDGs in these groups, we selected the proposed sample to be as large as possible while still spanning the full range of galaxy color available. The groups and the selected UDGs are shown in Figure~\ref{figure:galgrid_color}. Group A, containing UDGs 1A, 3A, 4A, and 5A\footnote{The UDG 2A observation was unsuccessful because the guide star acquisition failed.}, has a host halo mass $10^{12.5}\,{M}_{\odot}$ and $z=0.017$ (${\sim}75$ Mpc) \citep[][]{Yang2007}. UDGs 1-5B are located in group B with a host halo mass $10^{13.9}\,{M}_{\odot}$ and $z=0.0206$ (${\sim}90$ Mpc). These UDGs have $g$-band central surface brightnesses ranging from $23.8-26$ mag arcsec$^{-2}$ and effective radii from $2.5-3.5$ kpc \citep[][]{Greco2018IlluminatingSurvey}. Properties of the UDGs, calculated assuming the group distances, are detailed in Table~\ref{tab:gals}. Stellar-mass-to-light ratios for UDGs and dwarfs are typically of order unity \citep[e.g.][]{Greco2018AField, Pandya2018ThePhotometry}, so we calculate the stellar masses assuming $(M_*/L_{V})/({M_*/L_V})_\odot = 1$. \changeref{This assumption is an oversimplification, but we do not currently have the data needed to better constrain $M_*/L_{V}$. We discuss possible problems with this assumption in Section~\ref{sec:discussion}.} We use the galaxy sizes presented in \cite{Greco2018IlluminatingSurvey}, which are measured with a single S{\'e}rsic fit. We calculate the UDG $V$-band magnitudes using the HSC photometry and the conversions in \cite{Jester2005}. 

\subsection{Hubble Space Telescope Data}
We obtained deep follow up of the UDGs with the Wide Field Channel (WFC) of the Advanced Camera for Surveys (ACS) on the {\it Hubble Space Telescope} (\emph{HST}) (PID 15277, PI J. Greco; \citet{Greco2017WeighingGap}). ACS has the small angular resolution (${\sim}0\farcs1$) 
necessary to reliably differentiate between point sources (e.g., GCs) and background galaxies. We observed each UDG with the F814W filter (one orbit split into two 2486 sec dithers) and the F606W filter (one orbit split into two 2624 sec dithers). We chose these filters because we can achieve comparable depth in each with a single orbit.

We calibrated the data using the default \texttt{calacs} pipeline. The images had significant cosmic ray (CR) contamination, so we used \texttt{LACosmicX}\footnote{\url{https://github.com/cmccully/lacosmicx}} \citep{VanDokkum2001} to perform an initial CR removal. \texttt{LACosmicX} uses Laplacian edge detection to identify CR contaminated pixels. We required a Laplacian-to-noise of $6\sigma$ to identify a contaminated pixel, a fractional detection limit of $0.5$ for flagging pixels bordering an identified CR, and we defined the contrast limit between a CR and an underlying object with a Laplacian-to-noise of $6\sigma$. We then used \texttt{DrizzlePac} to remove the remaining CRs. \texttt{DrizzlePac} flags CR contaminated pixels as those which are only bright in one dither. We required a CR threshold of $15 \sigma$ and a convolution kernel width of $3.5$ pix.

\section{Globular Cluster Candidate Detection} \label{sec:GCdet}

In this section, we describe our GC detection procedure. First, we will summarize our overall detection approach, and then we will describe and motivate the detailed choices that we made.

\subsection{Source Identification} \label{sec:srcid}
To maximize our sensitivity to GCs, we identify point sources in each galaxy using the sum of the F814W and F606W images. We use these source positions to perform aperture photometry on the individual F814W and F606W images. We subtract the background from the high signal-to-noise detection image using a masked, sigma-clipped median subtraction. We perform this subtraction using a series of steps. First, we use \texttt{Photutils} to create a source mask of the image \citep[][]{Bradley2019Astropy/photutils:V0.7.2}. We require that the sources included in this mask contain at least $3$ pixels above a signal-to-noise threshold of $10$. We mask a region around each source which extends to twice the source size. Next, we create the background map for the masked image by running a sigma-clipped median smoothing with box size of $10\times10$ pix$^2$.

We detect point sources on the background subtracted detection image using \texttt{Photutils}. We require that sources have at least $3$ pixels above a detection threshold of $5 \sigma$. We then measure source fluxes on the individual F814W and F606W images and account for background by correcting each source flux by the background level in a circular annulus of inner (outer) radius $6$ ($10$) pix. We measure magnitudes in apertures of radius $2$, $3.5$, and $4$ pixels with an initial magnitude zeropoint of $26.0$ mag. We correct the magnitudes for aperture effects using the mock GC tests in Section~\ref{sec:mockGCs}. 

\begin{figure*}
\gridline{\fig{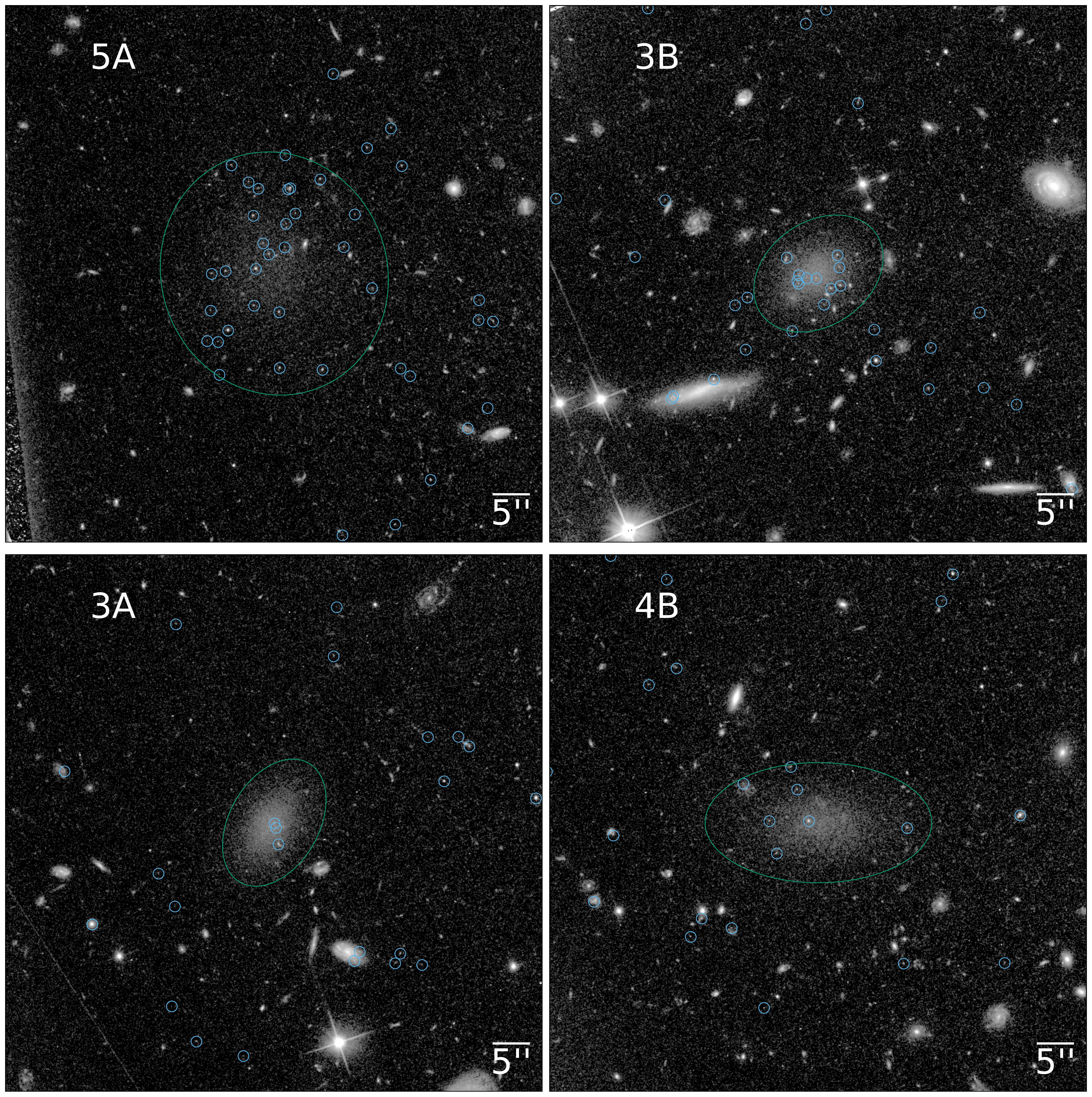}{\textwidth}{}}
\caption{ The GC candidates for UDGs 5A ({\it top left}), 3A ({\it bottom left}), 3B ({\it top right}), and 4B ({\it bottom right}), overlaid on the coadded F606W+F814W images. The UDGs in the top panels have significant GC detections, while those in the bottom panels do not. Each GC candidate, shown in blue, passes our optimized concentration and color cuts. The galaxy region is shown as a green ellipse which has the same size and shape as the galaxy, but extends to $2r_{\rm eff}$. Those candidates within the galaxy region are associated with the galaxy. At the distance of group A (B), $5''$ corresponds to ${\sim}1.8 (2.2)$ kpc \label{figure:GCgalgrid_606}}
\end{figure*}

Many of the recovered sources are background galaxies, residual CRs, and intra-group GCs. We first apply a cut on the source concentration, defined as $c=c_2-c_{3.5}$, where $c_i$ is the magnitude measured in the F606W band using an $i$-pixel radius aperture \citep[e.g.][]{Peng2011}. This cut removes both very extended and very localized objects, namely, extended background galaxies and CRs that contaminate a small number of pixels. Background galaxies at higher redshift can appear as point sources and cannot be removed with a concentration cut, but they have distinctive red optical colors and can be safely removed by applying a cut on \colorVIACS{}. We optimize these cuts using mock GC detection tests, described in Section~\ref{sec:mockGCs}. After this step, ${\sim}20\%$ of the original point sources within $2r_{\rm eff}$ of each UDG remain, and these remaining sources are shown in blue in Figure~\ref{figure:GCgalgrid_606} for UDGs 3A, 5A, 3B, and 4B.

After these cuts, point sources that are intra-group GCs, compact background galaxies, or even background fluctuations will remain. We will call these `background'. We subtract out this background contribution statistically using a measure of the average background level over the full field. First, we define GCs associated with the UDG as those in an elliptical aperture with semi-major axis $2 r_{\rm eff}$ centered on the UDG \citep[e.g.][]{Beasley2016}, as shown in green in Figure~\ref{figure:GCgalgrid_606}. We choose a region size of $2 r_{\rm eff}$ because we expect it to contain most of the GC population, but it is not so large that background contamination will be prohibitive. We find that variations on region size larger and smaller than $2 r_{\rm eff}$ do not significantly change our results. To estimate the background in this `galaxy region', we randomly fill each \emph{HST} image with apertures of the same size and shape as the galaxy region. We require that these `background regions' do not overlap with the galaxy region or with one another. We find the average number of sources in each background region that pass our concentration and color cuts, and define this as our background source density. In this average, we do not include those background regions that overlap with `bad' areas of the image, such as chip gaps or saturated stars. Before applying any completeness correction, we find ${\sim}0.3$ background sources/arcsec, corresponding to ${\sim}3$  background sources within $2r_{\rm eff}$ of the typical UDG.

\subsection{Mock Globular Cluster Tests} \label{sec:mockGCs}

To optimize the aforementioned color and concentration cuts and to assess our completeness, we injected mock GCs in 30 random locations in a circular region of radius $5 r_{\rm eff}$ around each UDG. We drew the GC magnitudes from the typical observed GC luminosity function (GCLF) modeled as a Gaussian with mean $M_{V}=-7.3$ mag and $\sigma=1.1$ mag \citep[][]{Miller2007}. We converted the $V$-band magnitude to F606W assuming $m_{V} - m_{V_{F606W}} = 0.13$ mag. We drew the GC \colorVI{} from a uniform distribution over the range of expected GC colors \change{$[0.3, 1.3]$} mags. Since $m_{\rm I} = m_{\rm I_{F814W}}+{\sim}{0.01}$ mag, this color range corresponds to $\colorVIACS \in [0.17, 1.17]$ mags. We created the mock GCs using \texttt{TinyTim} PSFs \citep{Krist201120Tim}, normalized and scaled by $10^{-0.4 \times (m-ZP)}$, where $m$ is the mock GC magnitude. We retrieved the zeropoints $ZP({\rm F606W})$ and $ZP({\rm F814W})$ from the \emph{HST} ACS website for the observation dates 2019-03-20, 2019-07-02, 2019-07-03, 2019-05-14, 2019-03-24, 2019-05-16, and 2019-07-04. The zeropoints were $ZP({\rm F606W}) = 26.50$ mag, and $ZP({\rm F814W}) = 25.94$ mag for all of the observations. We created 50 GC-enriched images by repeating this injection process.
\begin{figure*}
\gridline{\fig{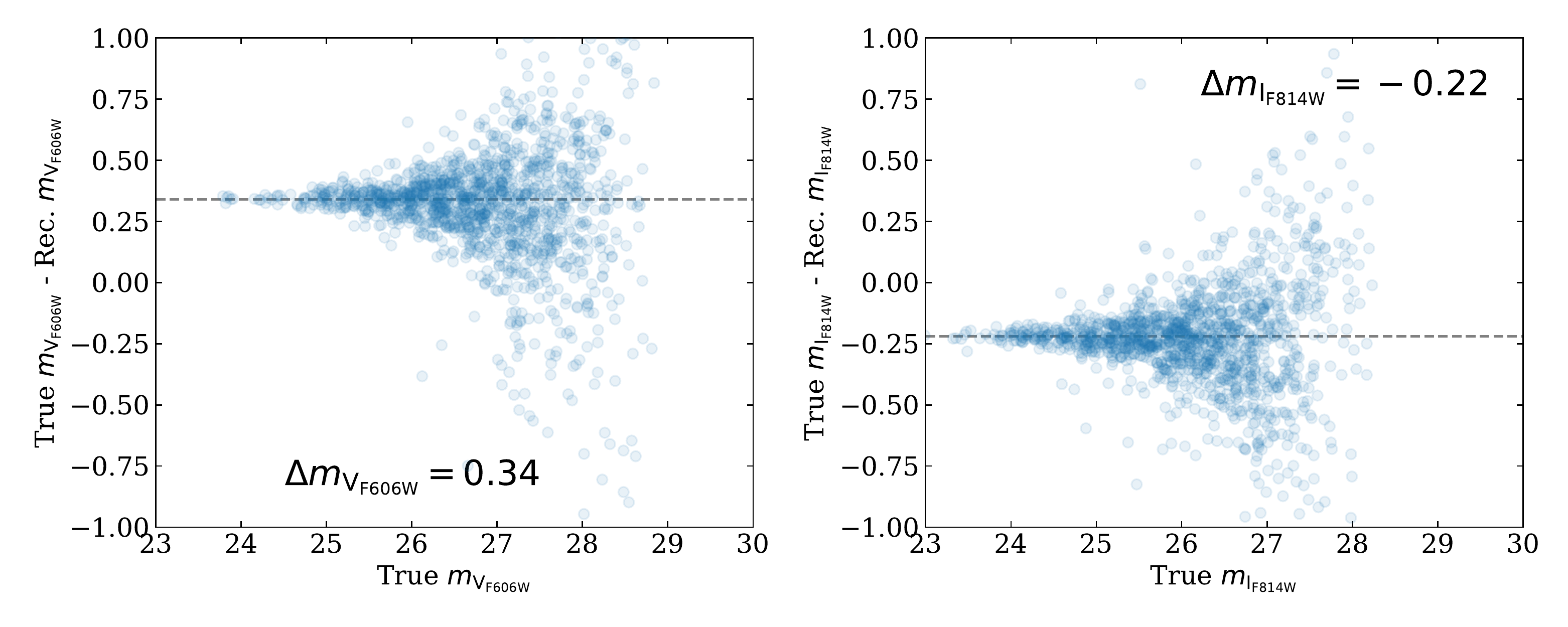}{0.9\textwidth}{}}
\caption{ The scatter in our recovered magnitude for UDG 5A, assuming an initial magnitude zeropoint of 26.0. Blue scatter points shows the magnitude difference between the recovered mock GC magnitudes and the true, injected mock GC magnitudes as a function of the true magnitude. The plot on the {\it left} shows the magnitude error for the F606W band, while the plot on the {\it right} shows the same for the F814W band. We find a magnitude zeropoint shift of 0.34 mag for F606W filter and -0.22 mag for the F814W filter. \label{figure:MagZero}}
\end{figure*}
\begin{figure*}
\gridline{\fig{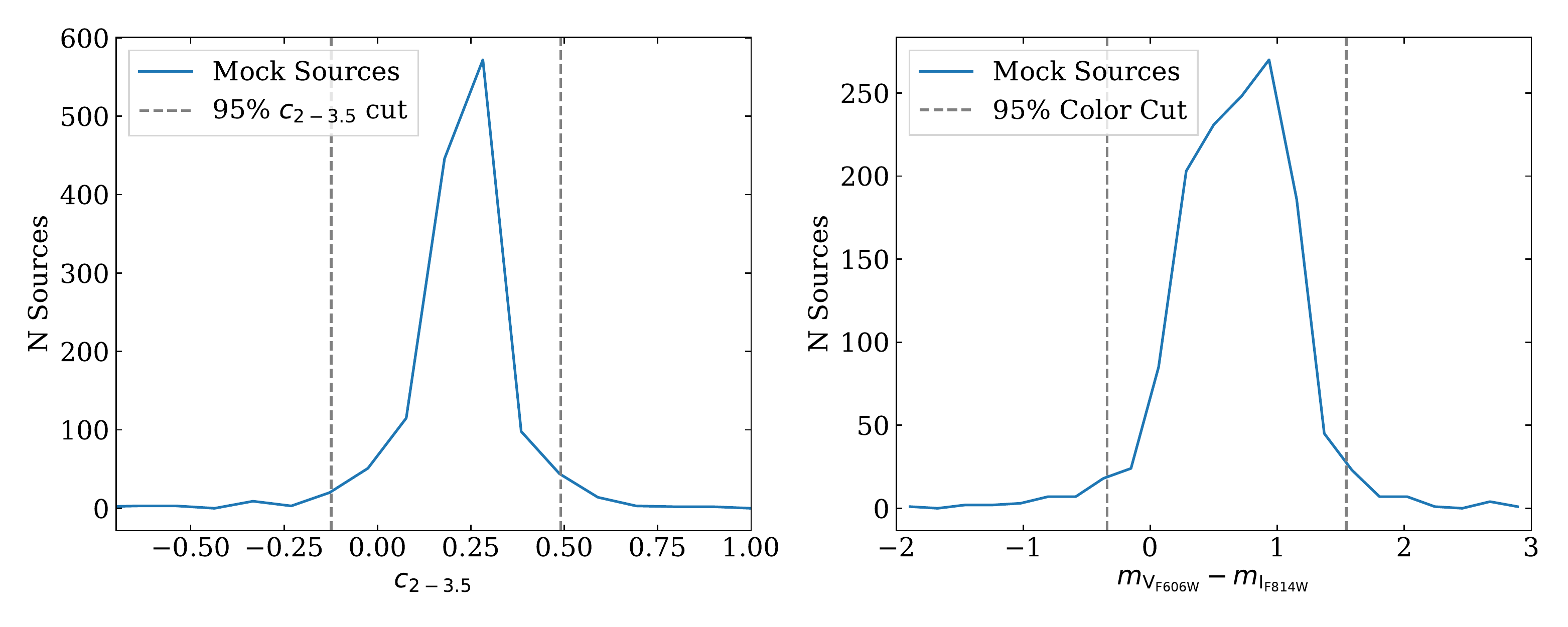}{0.9\textwidth}{}}
\caption{ {\it left} A histogram of the UDG 5A injected mock GC concentration $c_{2-3.5}$, defined as the magnitude measured in the F606W band with a 2 pixel radius aperture subtracted from that measured with a 3.5 pixel aperture. Grey dashed lines show our 95\% containment cuts. {\it right} A histogram of the recovered color \colorVIACS{} for our injected mock GCs in UDG 5A. Grey dashed lines show our 95\% containment cuts. \label{figure:cutOpt}}
\end{figure*}

We ran \texttt{Photutils} to detect GC candidates in each GC-enriched image following the same procedure as for our fiducial analysis. We recovered ${\sim}75\%$ of the injected sources. We used the recovered mock GC magnitudes to correct the 4-pixel aperture magnitude zeropoint for, e.g., aperture effects. We subtracted the recovered magnitude from the true magnitude, shown for UDG 5A as a function of true $m_{V_{F606W}}$ in Figure~\ref{figure:MagZero}. The magnitude zeropoint correction is the average of this magnitude difference for the brightest recovered sources ($m_{V_{F606W}} \lesssim 26$). We found that all UDGs required $ZP({\rm F606W}) = 26.34$ mag, and $ZP({\rm F814W}) = 25.79$ mag. We use the scatter in the zeropoint correction to estimate a magnitude error of $\pm 0.15$. We also use the scatter in the recovered color to estimate a color error of $\pm 0.2$.

\begin{figure*}
\gridline{\fig{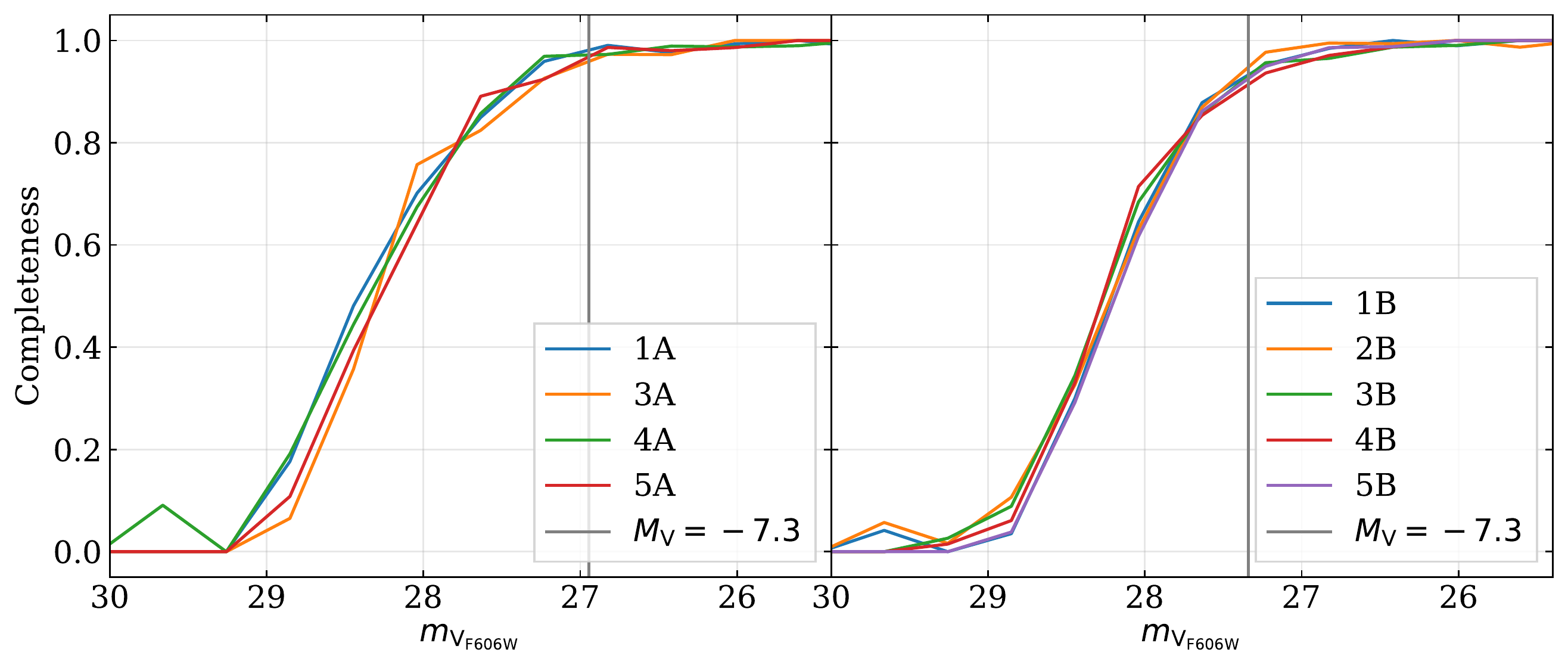}{0.9\textwidth}{}}
\caption{ Completeness curves. The completeness curve as a function of F606W magnitude for group A is shown on the {\it left} and that for group B is shown on the {\it right}. In all cases, we are complete to ${\sim}95\%$ at the GCLF peak $M_{V}=-7.3$, shown as a grey vertical line. \label{figure:completeness}}
\end{figure*}

We then optimized our color and concentration cuts. In Figure~\ref{figure:cutOpt}, we show histograms of the recovered mock GC color and $c_{2-3.5}$ for UDG 5A. We define our cut on each variable as the interval which contains 95\% of the injected GCs, shown as vertical lines in Figure~\ref{figure:cutOpt}. Since we apply a statistical background cut as described in Section~\ref{sec:srcid}, we aim to maximize completeness over purity with these cuts.  We list these cuts for each galaxy in Table~\ref{tab:GCs}. We also show the resulting completeness curves for each galaxy in Figure~\ref{figure:completeness}. Given these completeness curves and assuming that the GCLF has mean $M_{V}=-7.3$ and width $\sigma=1.1$ \citep[][]{Miller2007}, we expect to recover ${\sim}70-80\%$ of the GCs in our UDGs. Thus, we will correct our observations for completeness by multiplying the observed number of GCs by $1.2-1.4$. The exact completeness corrections for each UDG are reported in Table~\ref{tab:GCs}. \change{Note that we are not applying a spatial completeness correction. Previous work has suggested that such a correction would be a factor of ${\sim}2$ \citep[e.g.][]{VanDokkum2017, Saifollahi2020TheGalaxies}. Because of the uncertain nature of this correction, we choose to simply report the expected number of GCs within $2r_{\rm eff}$.}

\section{Results} \label{sec:results}
\subsection{Globular Cluster Detections} \label{sec:detections}

\begin{figure*}
\gridline{\fig{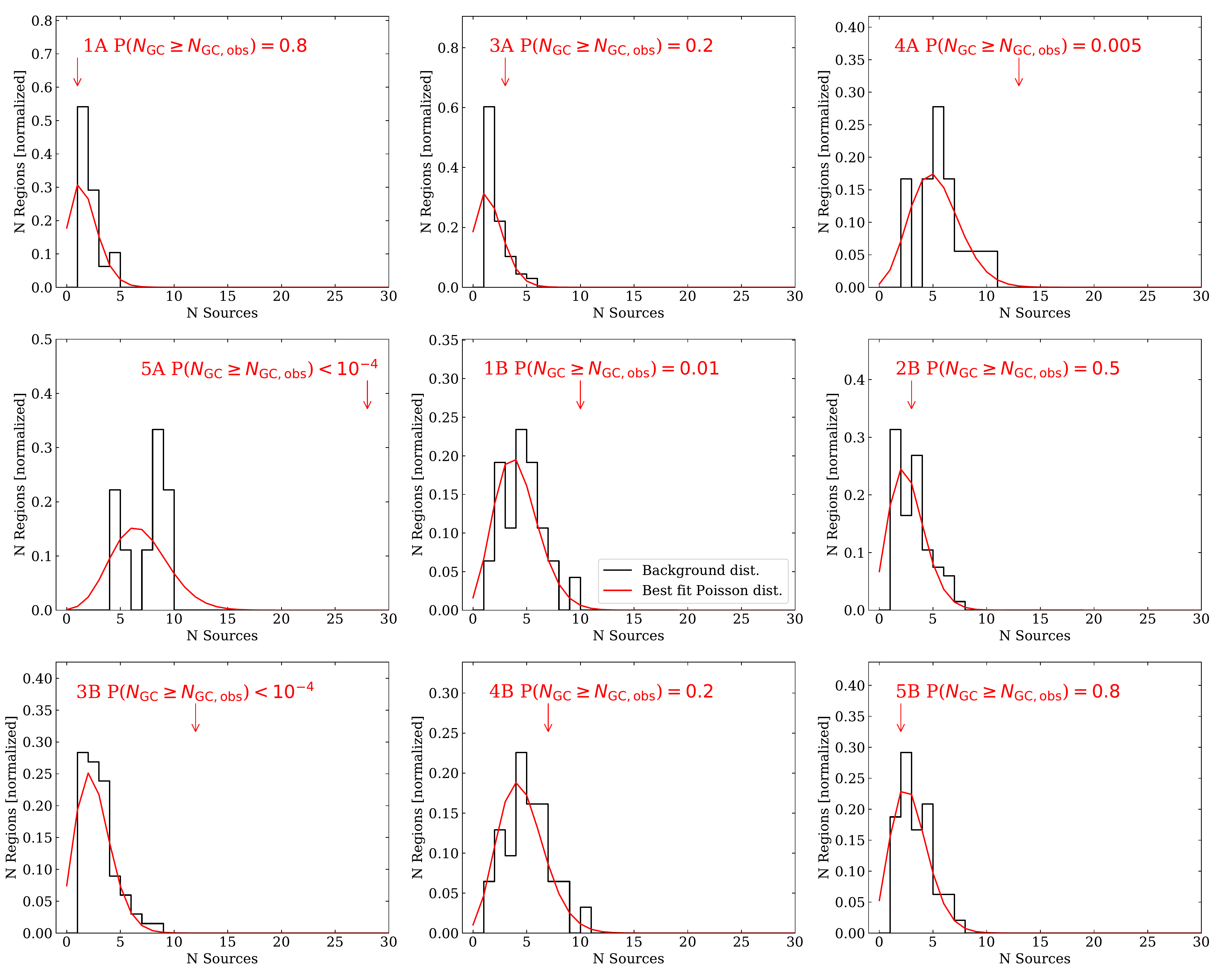}{\textwidth}{}}
\caption{ The number of GC candidates in the galaxy region compared to the GC candidate background for each UDG. The number of GC candidates detected in each galaxy region is shown as a red arrow. Histograms of the number of GC candidate detections in each background region are shown in black. Poisson fits to the background GC distributions are shown as red lines, and the resulting p-values for the GC detections are shown at the top of each panel. UDGs 5A and 3B show very significant detections with p-values smaller than $10^{-4}$. The rest of the UDGs have insignificant detections, with p-values $> 0.005$. \label{figure:GCbkg}}
\end{figure*}

In Figure~\ref{figure:GCbkg}, we show a histogram of the counts in the background regions for each UDG. We fit the background count distributions to Poisson distributions, which are shown in red. We use the Poisson distributions to assess the significance of an excess GC detection in each galaxy region. UDGs 5A and 3B show unambiguous globular cluster detections, with p-values smaller than $10^{-4}$. UDG 4A has a marginally significant p-value of $0.005$, so we do not consider it an unambiguous detection. In all other cases, we have non-detections. 

\begin{deluxetable*}{ccccccccc}
\tablenum{2}
\tablecaption{Ultra-Diffuse Galaxy Globular Cluster properties  \label{tab:GCs}}
\tablewidth{0pt}
\tablehead{
\colhead{Galaxy} & \colhead{$N_{\rm GC}$} & \colhead{$N_{\rm bkg}$} & \colhead{$S_{\rm N}$} & \colhead{$T$} & \colhead{\bigcell{c}{Completeness \\[-0.1cm] correction}} & \colhead{\bigcell{c}{$c_{2-3.5}$ \\[-0.1cm] cut}} & \colhead{\bigcell{c}{\colorVIACS{} \\[-0.1cm] cut}} & \colhead{$M_{\rm Halo,\,GC}\,[10^{10} {M}_{\odot}]$}}
\startdata
UDG-1A & $-1 \pm 2$ & $2.1 \pm 0.2$  & $-1 \pm 2$ & $-10 \pm 20$ & 1.2 & $[-0.13, 0.49]$ & $[-0.19, 1.45]$ & $[0,0.4]$ \cr \\ 
UDG-3A & $2 \pm 3$ & $2.0 \pm 0.2$ & $3 \pm 5$ & $30 \pm 60$ & 1.2 & $[-0.15, 0.48]$ & $[-0.3,1.65]$ & $[0,2]$ \cr \\
UDG-4A & $9 \pm 5$ & $6.3 \pm 0.7$ & $13 \pm 7$ & $150 \pm 80$ & 1.2 & $[-0.21, 0.49]$ & $[-0.34,1.6]$ & $[2,8]$ \cr \\
UDG-5A & $26 \pm 7$ & $8.4 \pm 1.1$ & $50 \pm 10$ & $600 \pm 200$ & 1.2 & $[-0.12, 0.49]$ & $[-0.34,1.54]$ & $[11,21]$ \cr \\
\hline
UDG-1B &  $8 \pm 5$ & $5.7 \pm 0.4$ &  $7 \pm 4$ & $80 \pm 50$ & 1.4 & $[-0.49, 0.59]$ & $[-0.43,1.98]$ & $[1,8]$ \cr \\ 
UDG-2B & $0 \pm 3$ & $3.7 \pm 0.3$ & $1 \pm 4$ & $6 \pm 50$ & 1.4 & $[-0.3, 0.54]$ & $[-0.51,1.69]$ & $[0,1.7]$ \cr \\ 
UDG-3B & $13 \pm 5$ & $3.6 \pm 0.3$ & $17 \pm 7$ & $200 \pm 90$ & 1.4 & $[-0.46, 0.56]$ & $[-0.42,1.77]$ & $[4,11]$ \cr \\
UDG-4B &  $3 \pm 4$ & $6.3 \pm 0.5$ &  $4 \pm 5$ & $50 \pm 60$    & 1.4 & $[-0.47, 0.53]$ & $[-0.44,1.93]$ & $[0,4]$ \cr \\
UDG-5B & $-1 \pm 3$ & $4.1 \pm 0.3$ & $-10 \pm 10$ & $-70 \pm 150$ & 1.4 & $[-0.33, 0.49]$ & $[-0.34,1.76]$ & $[0,0.6]$ \cr \\
\enddata
\tablecomments{$N_{\rm GC}$, $S_{\rm N}$, and $T$ have been background subtracted and completeness corrected, assuming a Gaussian GCLF with mean $M_{V}=-7.3$ and $\sigma = 1.1$. Error bars are entirely statistical. Negative values of $N_{\rm GC}$ are set to $0$. We show the lower and upper bounds on $M_{\rm Halo,\,GC}$, calculated using the calibration from \cite{Harris2017}.}
\end{deluxetable*}

We find the number of GCs in each UDG by subtracting the average background count from the number of galaxy GC candidates and correcting for completeness. We find that UDG 5A has $N_{\rm GC} = 26 \pm 7$, and UDG 3B has $N_{\rm GC} = 13 \pm 5$, where the reported uncertainties are entirely statistical. 
The remainder of our sample shows no statistically significant concentrations of GCs. The background subtracted and completeness corrected number of GCs detected in each galaxy is shown in Table~\ref{tab:GCs}. The completeness corrected average background for each UDG is shown in the third column.

Because the number of GCs varies with stellar and halo mass, it is standard to consider the GC specific frequency, which is the number of GCs normalized by the galaxy luminosity, $S_{\rm N} = N_{\rm GC} 10^{0.4(M_{V} + 15)}$. We show the specific frequency for each galaxy in the fourth column of Table~\ref{tab:GCs}. Eight of our UDGs have $S_{\rm N} \lesssim 10$ within their uncertainties. The exception is UDG 5A with $S_{\rm N} = 50 \pm 10$.

The specific frequency is best for comparing galaxies with the same mass-to-light ratios, so we also report the $T$ parameter, which is the stellar mass normalized number of GCs, $T=N_{\rm GC}/(M_*/10^9\,M_{\rm \odot})$. As shown in the fifth column of Table~\ref{tab:GCs}, UDGs 5A ($T\sim600$) and 3B ($T\sim200$) are the only UDGs with $T$ significantly greater than zero.

There is a suggestion that the number of GCs correlates with halo mass \citep[e.g.][]{Blakeslee1997}. If this correlation holds for UDGs \citep[e.g.][]{Harris2017}, we can use our UDG GC populations to calculate GC-inferred halo masses. Using the recent calibration from \cite{Harris2017}, the halo mass $M_{\rm Halo,\,GC}$ is given in units of solar masses by
$$
    \log \left( \frac{N_{\rm GC}}{M_{\rm Halo,\,GC}} \right) = -8.56 - 0.11 \log \left( M_{\rm Halo,\,GC} \right).
$$

We show upper and lower bounds on our GC-inferred UDG halo masses $M_{\rm Halo,\,GC}$ in the last column of Table~\ref{tab:GCs}. UDGs 5A and 3B are consistent with GC-inferred halo masses larger than $10^{11}\,{M}_{\odot}$, which may be larger than expected given current predictions of the UDG stellar-halo mass relation \citep[e.g.][]{Behroozi2019UniverseMachine:010}. The remainder have $M_{\rm Halo,\,GC} \lesssim 8 \times 10^{10} \, {M}_{\odot}$.

\changeref{Finally, we consider the average colors of the globular clusters in those UDGs with significant detections. After correcting for background contamination, we find that UDG 5A has a GC color distribution with mean $\langle V-I \rangle = 0.54 \pm 0.05$ and standard deviation $\sigma_{(V-I)} = 0.13 \pm 0.1$. UDG 3B has a GC color distribution with mean $\langle V-I \rangle = 0.51 \pm 0.07$ and standard deviation $\sigma_{(V-I)} = 0.19 \pm 0.1$. In calculating the GC population parameters for each UDG, we assume a color uncertainty of $\delta(V-I)=0.2$ on each individual GC. The mean GC colors are blue, as seen in GC systems in dwarf elliptical galaxies \citep[][]{Lotz2004TheHalos, Georgiev2009}.
The colors are also consistent with the diffuse galaxy light for our sample, which is a trend that has been seen in other UDG samples \citep[e.g.][]{Beasley2016, VanDokkum2017}, albeit with small samples thus far. 
}

\subsection{Globular Cluster Luminosity Function} \label{sec:GCLF}

\begin{figure*}
\gridline{\fig{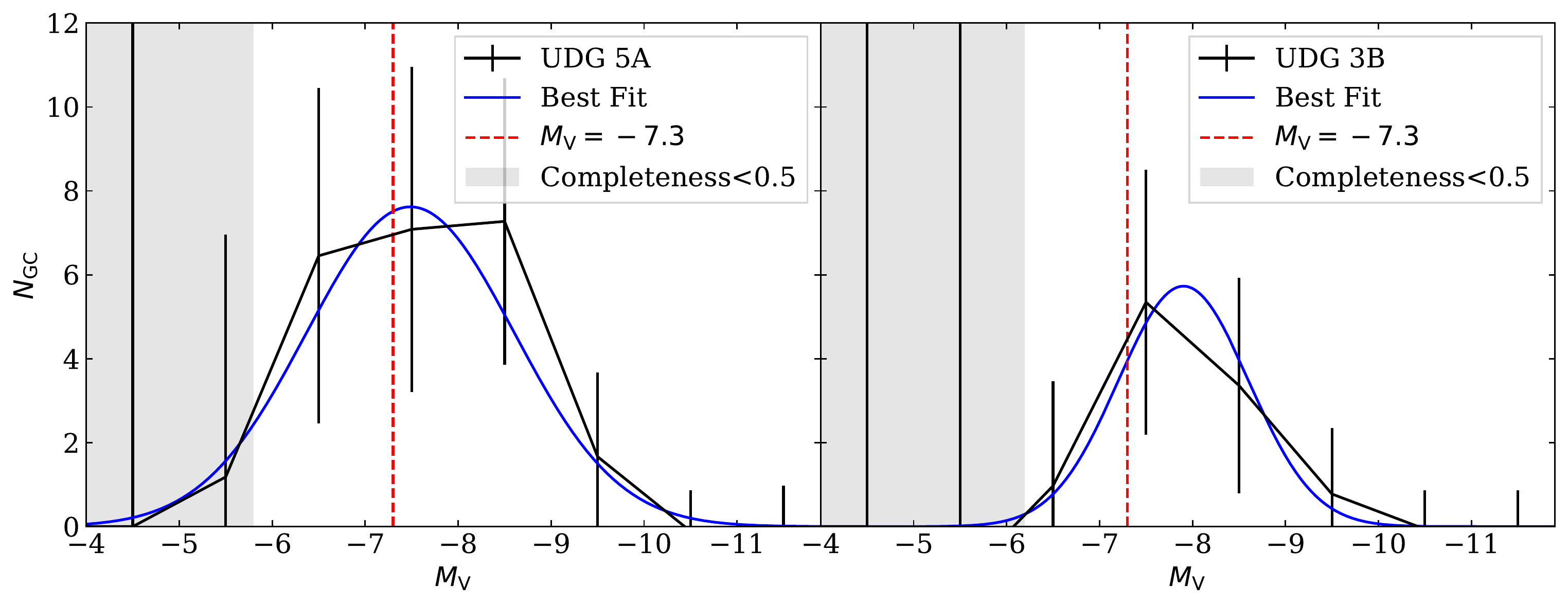}{\textwidth}{}}
\caption{ The globular cluster luminosity functions for UDG 5A {\it left} and 3B {\it right}. The shaded region has a completeness $<0.5$. A Gaussian fit to the unbinned data is shown in blue. We find that the GCLFs have means $\mu_{M_{V}} = -7.49 \pm 0.4, -7.90 \pm 0.5$, and widths $\sigma = 1.1 \pm 0.2, 0.7 \pm 0.4)$ for UDG 5A and 3B, respectively. The UDG GCLF peaks are both consistent with the expected $M_{V} = -7.3$ within ${\sim}2 \sigma$. \label{figure:GCLF}}
\end{figure*}

UDGs 5A and 3B both contain enough GCs to examine their GC luminosity distributions. In Figure~\ref{figure:GCLF}, we show in black the completeness-corrected GC luminosity functions for these two UDGs. To characterize the shape and peak of the luminosity functions, we fit them to Gaussian functions. We directly fit the unbinned data using a maximum likelihood formulation which is described in Appendix~\ref{sec:GCLFproc}. The resulting best fit signal Gaussian is shown in blue in Figure~\ref{figure:GCLF}. The shaded region of the plot has a completeness $<0.5$. UDG 5A was best fit by a Gaussian with mean $\mu_{M_{V}} = -7.49 \pm 0.4$ and width $\sigma = 1.1 \pm 0.2$. UDG 3B was best fit by a Gaussian with mean $\mu_{M_{V}} = -7.90 \pm 0.5$ and width $\sigma =0.7 \pm 0.4$. 


\begin{deluxetable*}{ccccccc}
\tablenum{3}
\tablecaption{Summary of comparison galaxy samples  \label{tab:comp}}
\tablewidth{0pt}
\tablehead{
\colhead{Type} & \colhead{Environment} & \colhead{$\langle \mu_{V} \rangle$ mag arcsec$^{-2}$} & \colhead{$M_*\,[{M}_{\odot}]$} & \colhead{$S_{\rm N}$} & \colhead{$M_{\rm Halo,\,GC}\,[{M}_{\odot}]$} & \colhead{Refs.}}
\startdata
\hline
Dwarf & Clusters, groups & $19-24$ & $10^{7.5-9}$ & $1$ & $10^{10-11}$ & \citealp{Miller2007, Harris2013} \\ 
\hline
UDGs & \bigcell{c}{High density \\ (clusters)} & $25-28$ & $10^{7-8}$ & $0-100$ & $10^{10-12}$ & \begin{tabular}{@{}c@{}}\citealp{Lim2018, Beasley2016} \\ \citealp{Peng2016, VanDokkum2017} \\ \citealp{Beasley2016a}\end{tabular} \\ 
\hline
UDGs & \bigcell{c}{Low density \\ (groups, isolated)} & $25-28$ & $10^{7-8}$ & $0-40$ & $10^{10-11}$ & \begin{tabular}{@{}c@{}}\citealp{Prole2019, Roman2019}\\ \citealp{VanDokkum2018,VanDokkum2018AMatter} \\ \citealp{Danieli2019AData}\end{tabular} \\ 
\enddata
\tablecomments{All numerical values are approximate. We assume a solar mass-to-light ratio.}
\end{deluxetable*}

\section{Observed trends in the globular cluster populations of ultra-diffuse galaxies}\label{sec:trends}
In the following, we will discuss the observed trends in UDGs and their GC populations. First, we will show that the GC luminosity functions of our UDGs with detections are consistent with observations of UDGs, dwarf galaxies, and high stellar mass galaxies (e.g., $L_*$ galaxies). Then, we will show that the GC abundances in our UDGs are consistent with both dwarf galaxies and other UDGs. Finally, we will perform a covariance analysis to better understand the relationship between galaxy stellar mass, size, and GC abundance.

We will compare our observations to previous results as detailed in Table~\ref{tab:comp}. These comparison samples contain dwarf galaxies, UDGs in low density environments, and UDGs in high density environments. We quote effective $V$-band surface brightness $\langle \mu_{V} \rangle$, defined as the average surface brightness within the effective radius. For all comparison samples, we report stellar masses assuming a mass-to-light ratio $(M_*/L_{V})/{ (M_*/L_{V})}_{\odot} = 1$. We will only compare to the galaxies with $10^7 < M_*/{M}_{\odot} < 10^9$ in order to match our UDG sample. 

We note that the dwarf sample from \cite{Miller2007}, the UDG samples from \cite{Lim2018} and \cite{VanDokkum2017} are among the least biased in $N_{\rm GC}$. The galaxies in these samples were not selected for analysis because they had large GC populations. We will focus on these works in Section~\ref{sec:covar}. 

\subsection{The UDG GCLFs are normal}
The peak of the globular cluster luminosity function does not appear to vary significantly between galaxies. \cite{Miller2007} found that the GCLF of Virgo dwarf elliptical galaxies is well modelled by a Gaussian with mean $M_{V} = -7.3 \pm 0.1$ and width $\sigma = 1.2 \pm 0.2$. They found that the GCLF peak is ${\sim}0.3$ mag brighter in giant spirals and ellipticals, but the variation in GCLF peak is typically smaller than $0.3$ mag within a single galaxy type.

Most studies of the UDG GCLF report peaks consistent with $M_{V}\sim -7.3$ \citep[e.g.][]{Peng2016,Roman2019,VanDokkum2017}. They typically measure widths $\sigma \sim 0.8$, which is slightly narrower than those measured in dwarf ellipticals. The GCLFs for the NGC 1052 UDGs DF2 and DF4 are abnormally bright with peaks at $M_{V} \sim -9$ \citep[][]{VanDokkum2018, VanDokkum2018AMatter}. These UDGs are also interesting because they may have small dark matter halos \citep[$M_{\rm halo} \lesssim 10^{8}\,{M}_{\odot}$;][]{Trujillo2019, Martin2018CurrentMatter, Danieli2019AData}. 

The GCLF peaks in UDG 5A ($\mu_{M_{V}} \sim -7.49 \pm 0.4$) and UDG 3B ($\mu_{M_{V}} = -7.90 \pm 0.5$) are both consistent with \cite{Miller2007}, as well as previous work on UDGs within ${\sim}2\sigma$. 
Their GCLF widths are also consistent with previous UDG observations ($\sigma \sim 0.8$) and with the \cite{Miller2007} result. In Section~\ref{sec:discussion}, we will use this similarity between GCLFs in UDGs and normal dwarf and high stellar mass galaxies to argue that these galaxy types share an early GC formation history.

\subsection{GC Abundance Correlations}
\begin{figure*}
\gridline{\fig{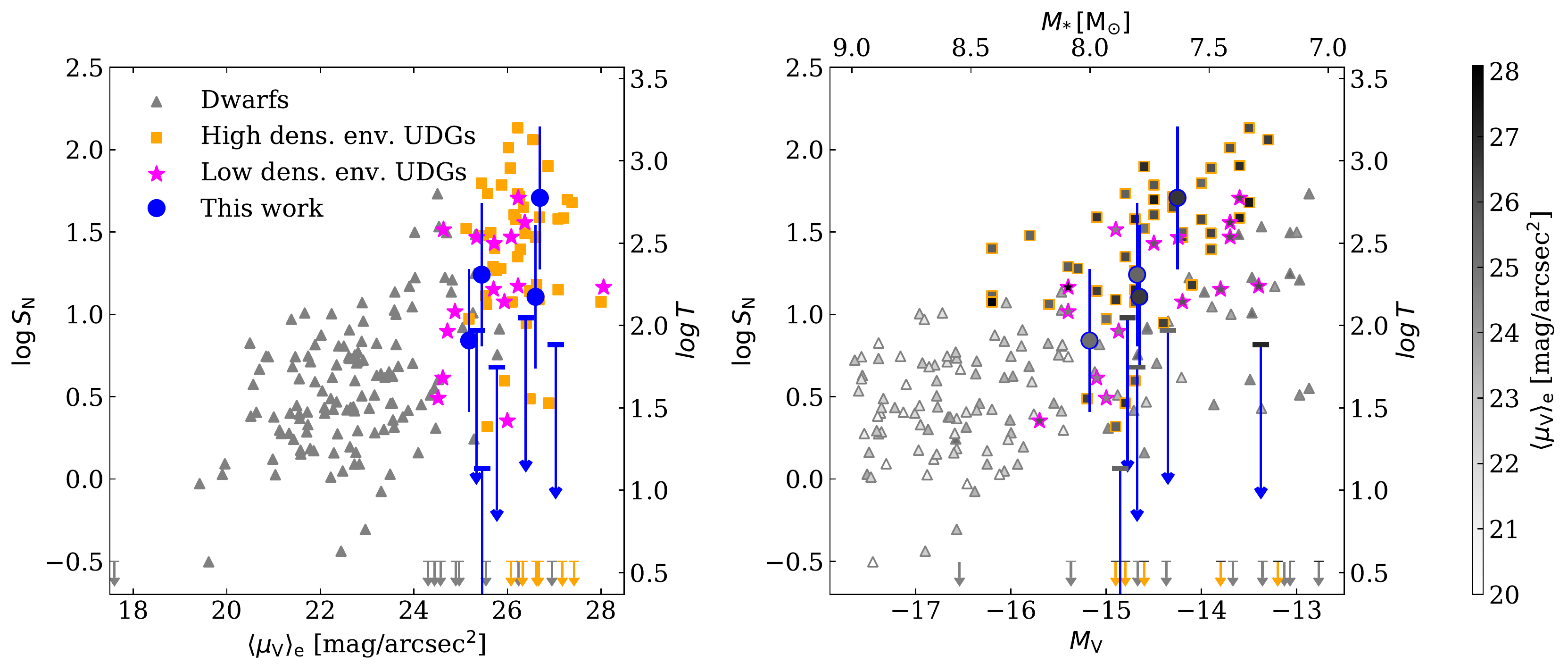}{\textwidth}{}}
\caption{ Properties of the UDG GC populations. The {\it left} panel shows the GC specific frequency (left axis) and $T$ parameter (right axis) as a function of effective $V$-band surface brightness. Our UDG sample is shown as blue circles for those UDGs with detections or possible detections and upper bounds for those without detections. High density environment UDGs from \citet{Beasley2016}, \citet{VanDokkum2017}, \citet{Peng2016}, \citet{Lim2018}, and \citet{Beasley2016a} are shown as orange squares, while low density environment UDGs from \cite{Roman2019, Prole2019,VanDokkum2018,VanDokkum2018AMatter} are shown as pink stars. Dwarfs from \citet{Miller2007} and \citet{Harris2013} are shown as grey triangles. The {\it right} panel shows GC specific frequency (left axis) and $T$ parameter (right axis) as a function of $M_{V}$ (bottom axis) and stellar mass $M_*$ (top axis), assuming a solar mass-to-light ratio \citep[e.g.][]{Greco2018AField, Pandya2018ThePhotometry}.  The shade of each marker denotes the surface brightness, with darker points having a lower surface brightness. Other than UDGs 5A and 3B, all of our UDGs are more consistent with the dwarf population and low density environment UDGs than the high density environment UDG population. \label{figure:GCprops}}
\end{figure*}
In Figure~\ref{figure:GCprops}, we compare our UDG GC specific frequencies and $T=N_{\rm GC}/(M_*/10^9)$ parameters, shown as large blue circles, to the dwarf galaxies (grey triangles), high density environment UDGs (orange squares), and UDGs in low density (group and isolated) environments (pink stars). As we mentioned, we only compare to those objects with $10^7 < M_*/{M}_{\odot} < 10^9$. 

First, we consider the population of dwarfs and UDGs as a whole. In the left panel of Figure~\ref{figure:GCprops}, we show the GC specific frequency and $T$ parameter as a function of surface brightness. Those objects with non-zero specific frequencies show a possible trend of increasing specific frequency with lower surface brightness, which has been discussed by \cite{Miller2007} for dwarf galaxies and \citet{Lim2018} and \citet{Forbes2020GlobularUDG} for UDGs. We choose to not directly analyze this correlation because both specific frequency and surface brightness directly depend on galaxy luminosity, which may lead to misleading correlations. Instead, we will recast it into the relationship between $N_{\rm GC}$, $r_{\rm eff}$, and $M_*$ in Section~\ref{sec:covar}.

The right panel of Figure~\ref{figure:GCprops} shows the specific frequency and $T$ parameter as a function of $M_{V}$ and $M_*$. Each data point is shaded according to its surface brightness. Like most UDGs with analyzed GC populations, the majority of our UDGs have specific frequencies more consistent with dwarf galaxies and low density environment UDGs than high density environment UDGs. UDG 3B has a slightly elevated specific frequency but is consistent with both the low density environment (pink stars) UDG population and dwarf (grey triangles) population within errors. UDG 5A shows a significantly elevated specific frequency more consistent with the high density environment (orange squares) UDGs than either the other low density environment UDGs or the dwarfs.

\begin{figure*}
\gridline{\fig{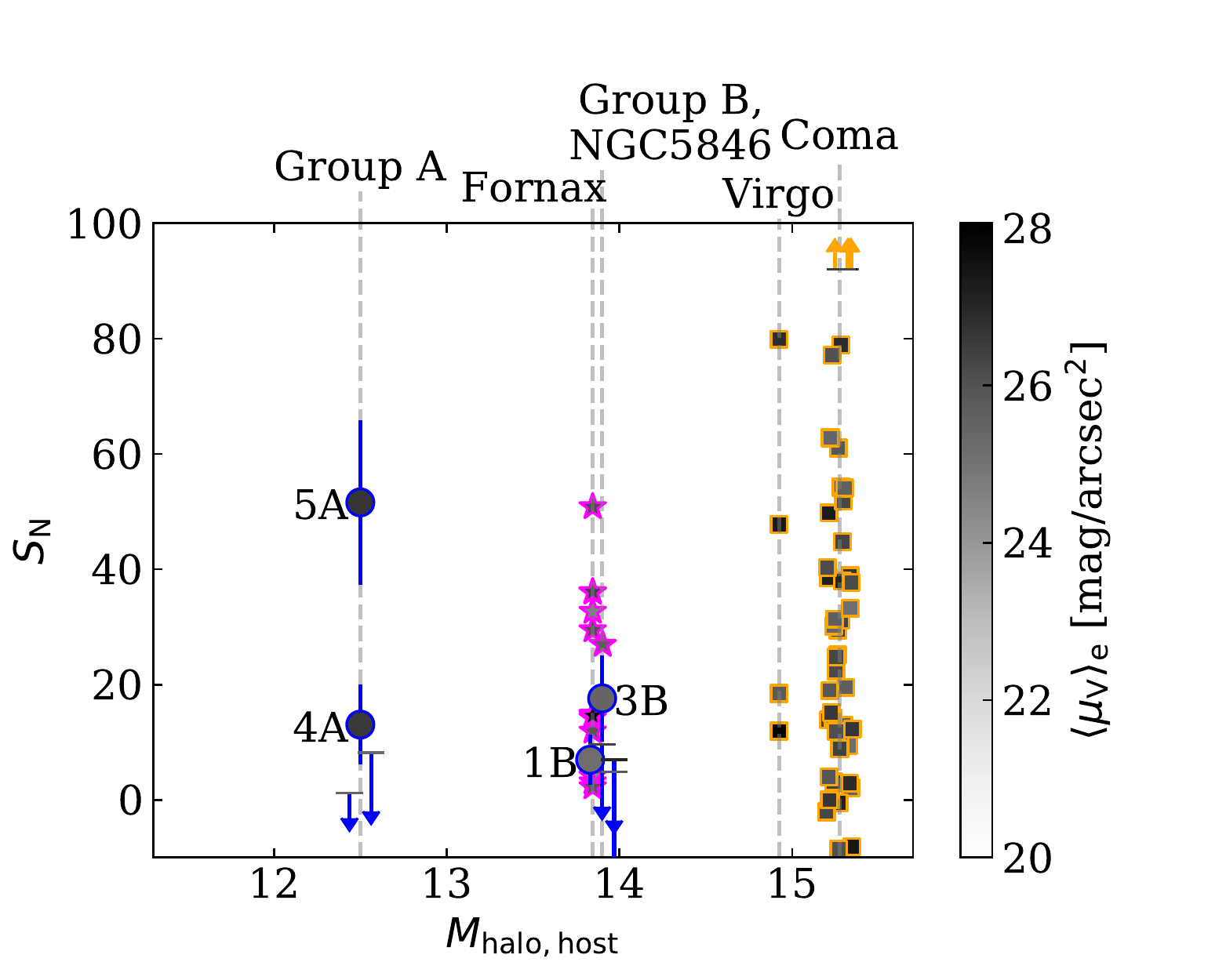}{0.7\textwidth}{}}
\caption{ A comparison of GC populations for UDGs in different environments. We show the specific frequency as a function of host halo mass. The results for our group UDGs are shown as blue circles at $M_{\rm halo, host}=10^{12.5}\,{M}_\odot$ (group A) and $M_{\rm halo, host}=10^{13.7}\,{M}_\odot$. Arrows show upper bounds for those UDGs with no detection. UDGs in high density, cluster environments \citep[][]{Lim2018, Beasley2016a, Beasley2016,Peng2016, VanDokkum2017} are shown as orange squares. UDGs in low density environments \citep[][]{Prole2019,Roman2019,VanDokkum2018AMatter,VanDokkum2018} are shown as pink stars. Our group UDGs appear to have smaller $S_{\rm N}$ than the high density environment UDGs and the scatter in our UDG $S_{\rm N}$ also appears to be smaller than that of the high density environment UDGs, although we require a larger sample size to determine if this environmental dependence is statistically significant. However, the other UDGs outside of clusters appear consistent with our results. \label{figure:halomass}}
\end{figure*}

From the decrease in UDG GC specific frequency from high density (orange squares) environments to low density environments (pink stars), we see that there may be a correlation between UDG environment and GC abundance. We highlight this possible trend in Figure~\ref{figure:halomass}, which shows the UDG GC specific frequency as a function of host halo mass for those objects where host halo masses are available. The richest GC systems ($S_{\rm N}\gtrsim 70$) are still all seen in clusters. This trend with environment is tantalizing. However, given the relation between halo mass and number of subhalos, observations of a $10^{15}\,M_{\odot}$ cluster can sample the dwarf satellite population much better than observations of a few low-mass groups can. It is possible that we simply require a larger sample of UDGs in low density environment to reveal the rare, extremely GC-rich systems. 

To naively estimate the required observations to test this environmental trend, note that ${\sim}10\%$ of the Coma UDGs reported in \cite{VanDokkum2017} and \cite{Lim2018} which lie in our mass range are on the extreme high tail of GC abundance, which we define as $S_{\rm N}\gtrsim 100$. Due to low statistics in the extreme GC abundance regime, there are large uncertainties on this estimate of $10\%$, so we will instead consider the range $5-15\%$. According to Poisson statistics, we require observations of ${\sim}30-80$ lower density environment UDGs to exclude with $2\sigma$ confidence the presence of UDGs with $S_N\gtrsim100$. If we expect a given group to host $5$ UDGs, we require observations of at least $6-16$ groups. While this is a very crude estimate, it highlights that more high-resolution observations of UDGs in groups would be powerful. Of course, the distribution of UDG halo mass may actually depend on environment, a possibility we discuss in Section~\ref{sec:covar}.

\section{The relationship between GC abundance, galaxy stellar mass, and galaxy size} \label{sec:covar}

\subsection{Covariance Analysis Methods}
\begin{figure*}
\gridline{\fig{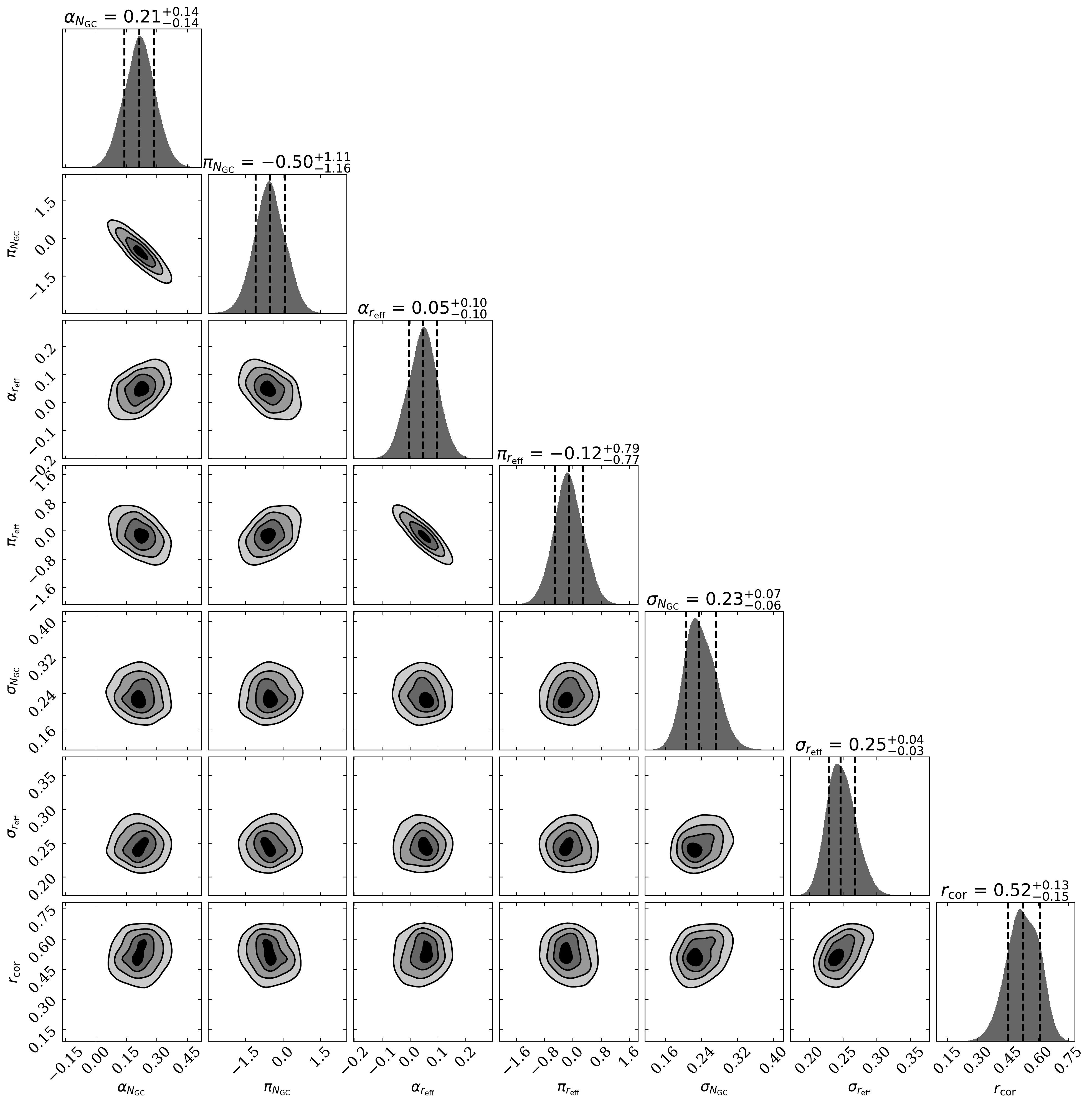}{0.6\textwidth}{},
          \fig{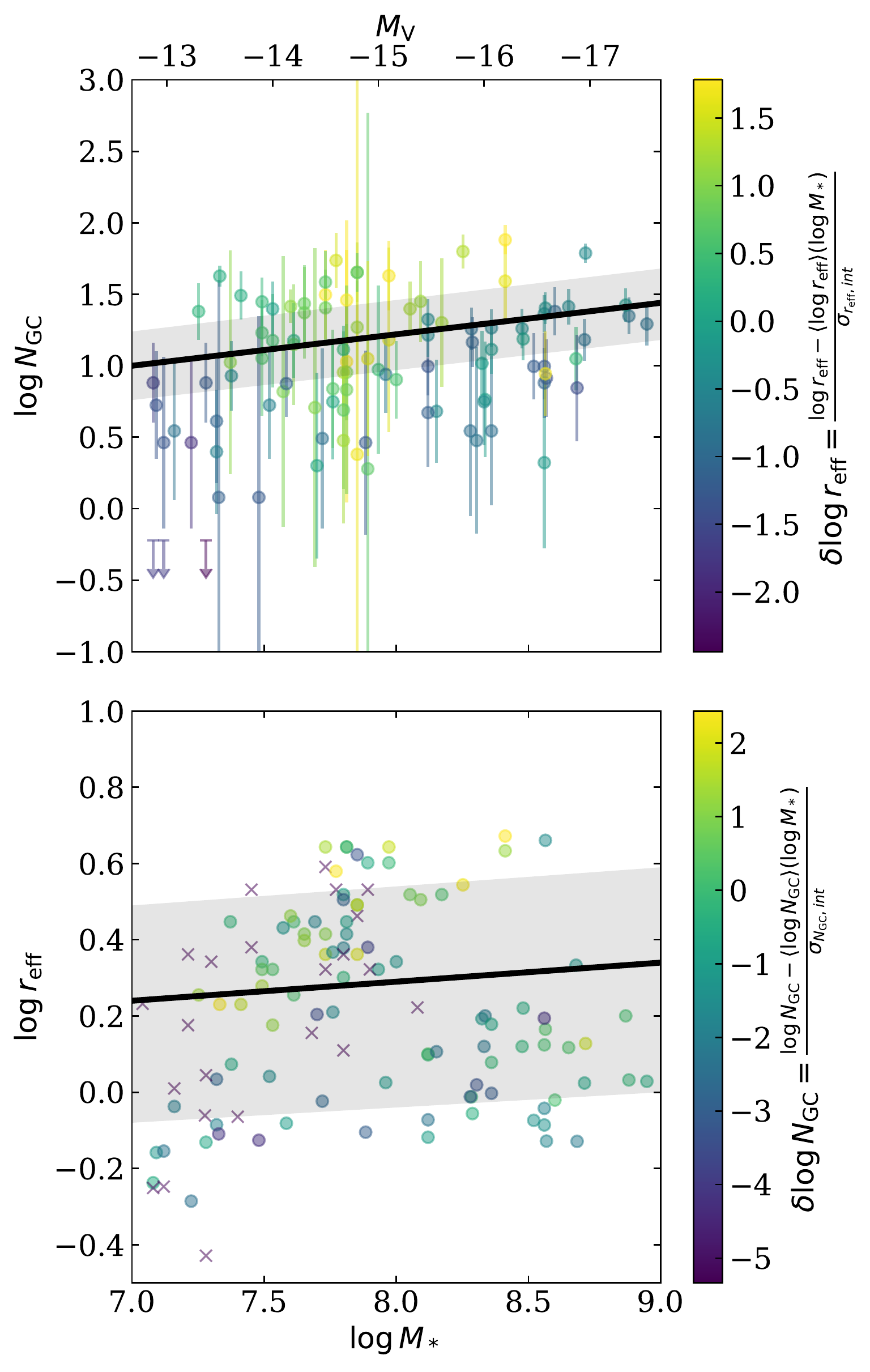}{0.4\textwidth}{}}
\caption{Posterior distribution for the covariance analysis of the relationship between stellar mass, effective radius, and globular cluster population for a sample of low mass galaxies. Relevant parameters are defined in Equation~\ref{eq:corr1} and Equation~\ref{eq:corr2}. Prior distributions are presented in Equation~\ref{eq:priors}. From the corner plot on the {\it left}, we see a possible linear dependence of $\log\,N_{\rm GC}$ on $\log\,M_*$, with large scatter. We do not detect any significant trend between $\log\,r_{\rm eff}$ and $\log\,M_*$. We find a significant correlation coefficient $r_{\rm cor}$. The dependence of $\log\,N_{\rm GC}$ on $\log\,M_*$ is shown on the {\it top right}. The corresponding magnitude scale is shown on the top axis, assuming $(M_*/L_{V})/(M_*/L_V)_{\odot}=1$. The data are shown as a scatter plot, with each point colored by the residual error in the best fit to its effective radius. Error bars on $\log\,N_{\rm GC}$ are shown, but the assumed errors on $\log\,M_*$ are suppressed for clarity. The best fit linear relationship between $N_{\rm GC}$ and $M_*$ is shown as a black line, and the intrinsic scatter is shown as a grey band. The dependence of $\log\,r_{\rm eff}$ on $\log\,M_*$ is shown on the {\it bottom right} in the same format as the $\log\,N_{\rm GC}$ vs $\log\,M_*$ plot, but with data points are colored by their residual $\log\,N_{\rm GC}$ error. Assumed errors on both $\log\,M_*$ and $\log\,r_{\rm eff}$ are suppressed for clarity. \label{figure:corner}}
\end{figure*}

In Figure~\ref{figure:GCprops}, we see a possible trend between surface brightness and GC specific frequency, suggesting a possible relationship between stellar mass (or equivalently, luminosity), galaxy size, and GC abundance. To gain further insight into this relationship, we performed a Bayesian fit to the UDG and dwarf samples. We assumed that the joint $r_{\rm eff}$ and $N_{\rm GC}$ distribution can be described by a multivariate Gaussian, where the mean of the Gaussian depends on stellar mass $M_*$. Specifically, the Gaussian is described by the mean vector and covariance matrix
\begin{gather} \label{eq:corr1}
    \boldsymbol{\mu} = \left[\,\alpha_{r_{\rm eff}} \log M_* + \pi_{r_{\rm eff}}, \alpha_{N_{\rm GC}} \log M_* + \pi_{N_{\rm GC}}\,\right] \\
    \label{eq:corr2}
    \boldsymbol{\Sigma} = \begin{pmatrix}
    \sigma_{r_{\rm eff}}^2 & r_{\rm cor} \times (\sigma_{r_{\rm eff}}\sigma_{N_{\rm GC}})  \\
    r_{\rm cor} \times (\sigma_{r_{\rm eff}}\sigma_{N_{\rm GC}}) & \sigma_{N_{\rm GC}}^2 \\
\end{pmatrix}
\end{gather}
Here, $r_{\rm cor} = {\rm Corr}(\log\,r_{\rm eff}, \log\,N_{\rm GC})$ is the correlation coefficient, distinct from the effective radius $r_{\rm eff}$. Moreover, we have defined
\begin{gather}
    \sigma_{r_{\rm eff}} = \sqrt{\sigma_{r_{\rm eff}, intr}^2 + \sigma_{r_{\rm eff}, obs}^2 + (\alpha_{\rm r_{eff}} \sigma_{M_*, obs})^2} \\
    \sigma_{N_{\rm GC}} = \sqrt{\sigma_{N_{\rm GC}, intr}^2 + \sigma_{N_{\rm GC}, obs}^2 + (\alpha_{\rm N_{GC}} \sigma_{M_*, obs})^2},
\end{gather}
where $\sigma_{i, obs}$ is the observational uncertainty for $i\in \{ {\rm \log N_{GC}, \log r_{eff}} \}$, and $\sigma_{i, intr}$ is the intrinsic scatter in the relationship between $i$ and $\log M_*$. The terms containing $\alpha_{\rm i} \sigma_{M_*, obs}$ account for the scatter resulting from the observational uncertainty of $M_*$. We assume $\sigma_{r_{\rm eff}, obs} = 0.1$ dex, and we calculate $\sigma_{N_{\rm GC}, obs}$ from the error reported for each $N_{\rm GC}$ measurement. We fit for both $\sigma_{r_{\rm eff}, intr}$ and $\sigma_{N_{\rm GC}, intr}$.

We fit this Gaussian to the dwarf sample from \cite{Miller2007}, the UDG samples from \cite{Lim2018} and \cite{VanDokkum2017}, and our UDG sample. 
A number of these objects have null or negative GC detections (negative detections in a given UDG arise from the statistical background subtraction). We do not input these values into our likelihood in the same way as positive detections; instead, before each likelihood evaluation, we resample these values from a Gaussian centered at zero and with a width corresponding to the reported error on that null detection. We cut off this Gaussian at zero and at $3\sigma_{\rm N_{GC}}$. The upper cutoff should not affect the results, and is to ensure that we do not randomly select an extremely large value of $N_{\rm GC}$, which is improbable, but would increase our computation time. 

We evaluate our likelihood using the \texttt{dynesty} sampler \citep[][]{Speagle2020Dynesty:Evidences}. We assume heaviside priors with the following limits:
\begin{equation} \label{eq:priors}
\begin{gathered}
    \alpha_{r_{\rm eff}} \in [-1,1],\,\pi_{r_{\rm eff}} \in [-3,3],\,\sigma_{r_{\rm eff}} \in (0,3], \\
    \alpha_{N_{\rm GC}} \in [-1,1],\,\pi_{N_{\rm GC}} \in [-3,3],\,\sigma_{N_{\rm GC}} \in (0,3], \\
    r_{\rm cor} \in [-1,1]
\end{gathered}
\end{equation}

\begin{figure}
\gridline{\fig{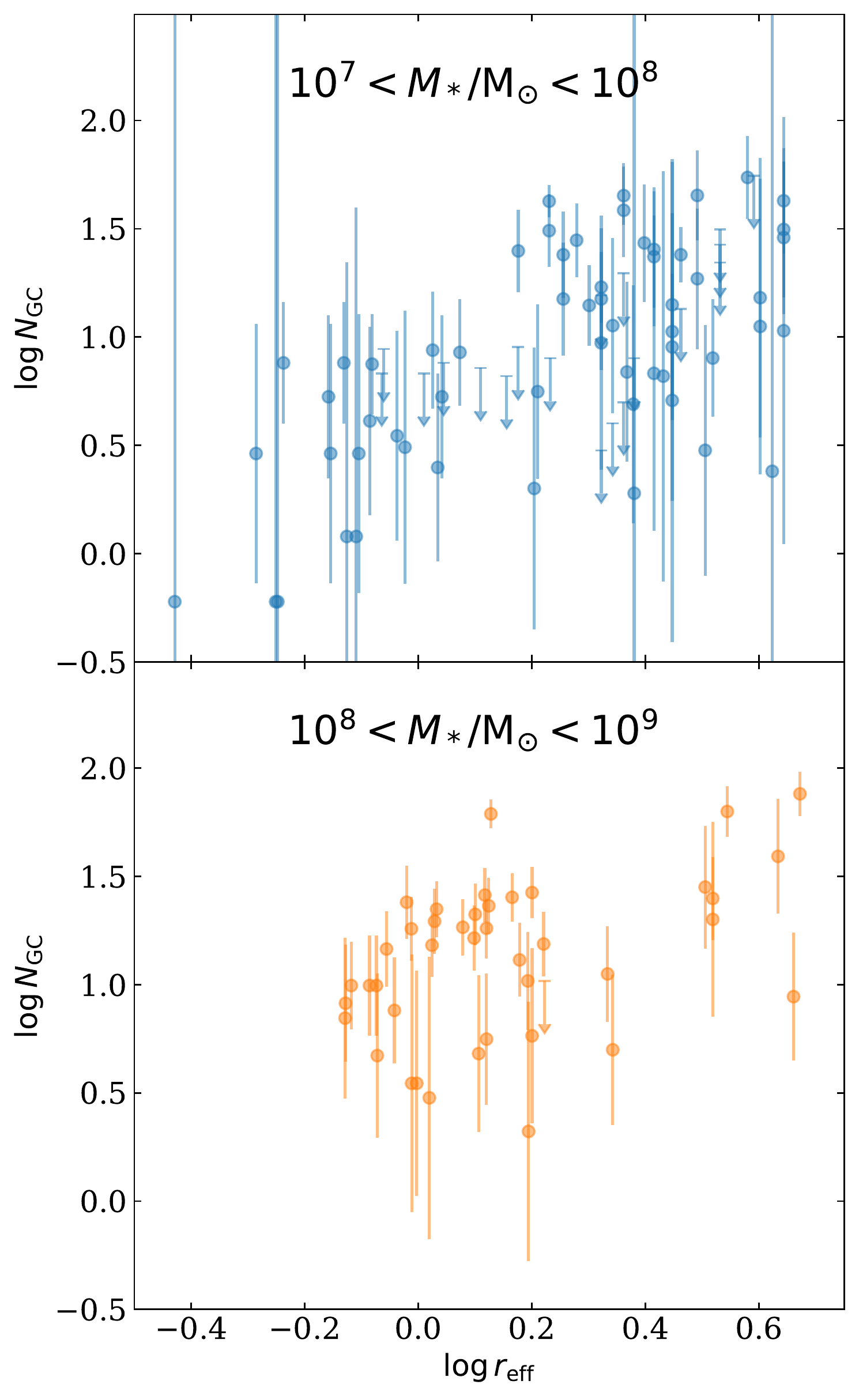}{0.4\textwidth}{}}
\caption{ Correlation between GC abundance and effective radius. The {\it top} panel shows the dependence of $\log\,N_{\rm GC}$ on $\log\,r_{\rm eff}$ for all UDGs and dwarf galaxies with low stellar masses $10^{7}<M_*/M_\odot<10^{8}$. Arrows denote the upper $2\sigma$ limit on $\log N_{\rm GC}$ for galaxies with null GC detections. The {\it bottom} panel shows the same for galaxies with higher stellar masses $10^{8}<M_*/M_\odot<10^{9}$. The positive correlation can be clearly seen for galaxies in the higher mass bin ({\it bottom} panel), and is marginally visible in the lower mass bin ({\it top} panel). \label{figure:NGCReffCorr}}
\end{figure}

We run \texttt{dynesty} until it reaches the stopping criteria of $\Delta \log Z=3$. 
In Figure~\ref{figure:corner}, we show the resulting posterior space and median-best fit mass-size and GC-stellar mass relations. We find a weak linear relationship 
\begin{equation}
    \log\,N_{\rm GC} = (0.21^{+0.14}_{-0.14}) \log\,M_* + (-0.50^{+1.11}_{-1.16}),
\end{equation} 
with intrinsic scatter $\sigma = 0.23^{+0.07}_{-0.06}$. We do not find a significant mass-size relationship 
\begin{equation*}
    \log\,r_{\rm eff} = (0.05^{+0.10}_{-0.10})\log\,M_* + (-0.12^{+0.79}_{-0.77}),
\end{equation*}
with intrinsic scatter, $\sigma=0.25^{+0.04}_{-0.03}$. We see a large correlation coefficient $r_{\rm cor}=0.52^{+0.13}_{-0.15}$ between $\log\,N_{\rm GC}$ and $\log\,r_{\rm eff}$. These relations and correlations can be seen in the left panel of Figure~\ref{figure:corner}, where we show $\log\,N_{\rm GC}$ vs. $\log\,M_*$ and $\log\,r_{\rm eff}$ vs. $\log\,M_*$. The color of each point in the upper (lower) panel corresponds to the residual error $\delta \log\,r_{\rm eff}$ ($\delta \log\,N_{\rm GC}$). There is an increase in $\delta \log\,N_{\rm GC}$ with larger $\log\,r_{\rm eff}$ and likewise in $\delta \log\,r_{\rm eff}$ with larger $\log\,N_{\rm GC}$.

\subsection{The $N_{\rm GC}-r_{\rm eff}$ correlation}
In Figure~\ref{figure:NGCReffCorr}, we show the relationship between $\log\,N_{\rm GC}$ and $\log\,r_{\rm eff}$ in two stellar mass bins, $10^7 < M_*/M_\odot < 10^8$ and $10^8 < M_*/M_\odot < 10^9$. The following discussion will explore possible origins of this relationship. We will first consider baryonic processes which could modify both galaxy size and GC abundance. Then, we will discuss an alternate scenario where galaxy size and GC abundance mutually depend on halo-related parameters. Throughout this discussion, we will assume that the high number of GCs per unit stellar mass in dwarf galaxies is evidence for an underlying relationship between GC abundance and halo mass \citep[e.g.][]{Harris2013}, although this relation may or may not hold \citep[e.g.][]{El-Badry2019ThePopulations}. Throughout, we also favor scenarios in which the formation of GCs in UDGs and higher surface brightness dwarfs are similar, given that their GCLFs are similar. 

Since we assume an $N_{\rm GC}-M_{\rm halo}$ relation at low stellar masses, the large scatter in UDG GC populations corresponds to a large scatter in UDG halo masses. This suggests either that the SHMR has large scatter in $M_{\rm halo}$ at fixed stellar mass, for low stellar masses, or that low stellar mass galaxies cannot be treated as a single, continuous population. Unless stated otherwise we will assume large scatter in the SHMR at fixed stellar mass. Because of the large scatter in the observed $N_{\rm GC}-M_*$ relation and in the assumed $N_{\rm GC}-M_{\rm halo}$ relation, there is room for large scatter in $M_{\rm halo}$ at fixed $M_*$ while preserving the halo/stellar mass-$N_{\rm GC}$ relations. Towards the end of our discussion, we will briefly mention an alternate scenario where the $M_{\rm halo} \lesssim 10^{11}\,{M}_{\odot}$ UDGs follow a standard SHMR, while some special physics forms the $M_{\rm halo} \gtrsim 10^{11}\,{M}_{\odot}$ UDGs.

\subsubsection{Baryonic Processes}
A possible explanation of the $N_{\rm GC}-r_{\rm eff}$ correlation is that the GC abundance and radius of a galaxy are both affected by the same baryonic process during the course of the galaxy's evolution. In particular, internal feedback can cause a galaxy to puff up \citep[][]{DiCintio2017}, and may also be able to increase the GC formation efficiency \citep[][]{Ma2020Self-consistentGalaxies}. Alternatively, mergers may disturb the baryonic content of galaxies in a way that can both increase galaxy size \citep[][]{Wright2020TheRomulus25}, and modify GC formation efficiency and survival rate \citep[][]{Kruijssen2014GlobularEvolution}. However, these mechanisms raise the question: does such baryonic physics affect the GCLF? It is observed that the shape and peak of the GCLF are similar in UDGs and dwarf ellipticals, and since it is thought that GC disruption shapes the GCLF \citep[e.g.][]{Kruijssen2014GlobularEvolution}, that disruption must operate in the same way in these galaxy types. It is difficult to think of a baryonic process which would increase a galaxy's size and GC abundance without modifying GC disruption and, consequently, the GCLF. Determining if such a mechanism exists will require further study, on varied feedback levels and merger histories on the GCLF, for example.

\subsubsection{Halo mass connection}
It may be easier to devise mechanisms that can generate the $N_{\rm GC}-r_{\rm eff}$ correlation without modifying the GCLF if we do not directly rely on complex baryonic processes. In particular, the GC abundance and galaxy size may mutually depend on properties of the galaxy dark matter halo. Such a mutual dependence does not contradict any of our observations because of our assumption of large scatter in the SHMR at low stellar mass. Specifically, given the large scatter in $M_{\rm halo}$ at fixed $M_*$, there could be a correlation between $M_{\rm halo}$ and $r_{\rm eff}$ but no observable correlation between $r_{\rm eff}$ and $M_*$. Similarly, the dependence of $N_{\rm GC}$ on a halo-related parameter would remain consistent with our observed $N_{\rm GC}-M_*$ relation. As a specific example of a mutual halo dependency, \cite{Kravtsov2013TheGalaxies} and \cite{Harris2013} have suggested that in the high stellar mass regime $r_{\rm eff}$ and $N_{\rm GC}$, respectively, may depend on the virial radius. It is possible that this mutual dependence could extrapolate to the low stellar mass regime.

Alternatively, we can remove the assumption of large scatter in the SHMR at low stellar masses, meaning that normal dwarfs can have a narrow range of halo mass at fixed stellar mass (e.g., 0.2 dex; \citealp{Garrison-Kimmel2016OrganizedGalaxies}), but UDGs have a different distribution in $M_{\rm halo}$. \cite{Forbes2020GlobularUDG} put forward a scenario like this, in which one population of UDGs could form with $M_{\rm halo} \lesssim 10^{11}\,{M}_{\odot}$ but large sizes ($r_{\rm eff}>1.5$ kpc) if they inhabit high angular momentum halos or experienced more extreme feedback \citep{Carleton2019TheHeating, Liao2019Ultra-diffuseSimulations, DiCintio2017, Tremmel2019TheSimulation}. On the other hand, a special class of UDGs with $M_{\rm halo} \gtrsim 10^{11}\,{M}_{\odot}$ could have formed as failed $L_*$ galaxies \citep{VanDokkum2017}. The UDGs with $M_{\rm halo} \lesssim 10^{11}\,{M}_{\odot}$ could be consistent with, e.g., an extrapolation of a standard SHMR and would have normal GC abundances given their stellar masses \citep[e.g.][]{Garrison-Kimmel2016OrganizedGalaxies}. In contrast, the $M_{\rm halo} \gtrsim 10^{11}\,{M}_{\odot}$ UDGs would have much lower stellar-to-halo mass ratios which would not satisfy a standard SHMR and would result in GC abundances which are abnormally high given the stellar masses of those UDGs.

\section{Summary and Future Work} \label{sec:discussion}
We have studied the GC populations of nine UDGs in group environments and with surface brightnesses $\langle \mu_V \rangle \approx 25-28$ mag arcsec$^{-2}$ and effective radii $r_{\rm eff} \sim 2.5-3.5$ kpc. We also have examined trends of GC populations in UDGs more broadly. Our main conclusions are:

\begin{itemize}
    \item The bulk (7/9) of our lower-density environment UDGs have GC abundances ($S_{\rm N} \lesssim 10$) consistent with normal dwarf ellipticals. 
    \item Two of our UDGs, UDGs 5A and 3B, have more GCs than expected for their stellar mass. They each have a GCLF with both peak and width consistent with the GCLFs in normal dwarf and high stellar mass galaxies and in other UDGs. 
    \item The most GC-rich ($S_{\rm N} \gtrsim 70$) UDGs so far have been found in the densest environments. Future observations of UDG GC populations in group environments are required to constrain the presence of such objects at lower densities.
    \item Combining well-defined UDG and dwarf elliptical samples, we see a possible positive dependence between $N_{\rm GC}$ and $M_*$ (slope $= 0.21^{+0.14}_{-0.14}$). We do not see a significant stellar mass-size relation (slope $= 0.05^{+0.10}_{-0.10}$).
    \item We find a positive correlation between $N_{\rm GC}$ and $r_{\rm eff}$ at fixed stellar mass ($r_{\rm cor} = 0.52^{+0.13}_{-0.15}$), which is similar to the surface brightness-$N_{\rm GC}$ relation that has been noted in the literature. Possible origins for this correlation include a mutual dependence of $N_{\rm GC}$ and $r_{\rm eff}$ on properties of the galaxy dark matter halo or a connection through the baryonic processes such as those involved in internal feedback or caused by galaxy mergers.
\end{itemize}

Further constraints on these observed trends, in particular on the possible environmental trend and the correlation between GC abundance and galaxy size, will require further observational and theoretical work. First, we would benefit from theoretical work modelling the effects of strong feedback, extreme variations in merger history, and enhanced/reduced star formation on GC formation. From the observational side, GC-independent halo mass constraints (e.g., dynamical measurements) would also provide a very powerful constraint on the discussed models. In particular, it would allow us to evaluate the validity of the $N_{\rm GC}-M_{\rm halo}$ relation in the UDG regime, which has been called into question by some theoretical work \citep[e.g.][]{El-Badry2019ThePopulations}.

Moreover, we require a better understanding of the space density of galaxies as a function of surface brightness. The low stellar mass galaxy samples which we consider have unknown number densities, and they are not volume complete, which makes it difficult to define their SHMR. This incompleteness could also lead to an artificially flat mass-size relation, although \cite{Danieli2019RevisitingGalaxies} also found no mass-size relation in their volume-complete sample of Coma cluster galaxies. 

\change{We have not accounted for the uncertainties in the distances to our UDGs, which could modify our constraints on $N_{\rm GC}$. In future work, \cite{GrecoXX} will measure the redshift distribution of the \cite{Greco2018IlluminatingSurvey} sample, enabling statistical distance corrections.}

Our results are also subject to the assumption of a constant mass-to-light ratio $(M_*/L_V)/(M_*/L_V)_{\odot} = 1$. Our covariance analysis results could be explained if, for example, there is a systematic trend in mass-to-light ratio with size. In the future, a more rigorous determination of stellar mass is necessary to assess the true relationship between GC abundance, galaxy stellar mass, and galaxy size. For example, multi-band SED fitting or spectroscopic fitting in the LSB regime, although difficult, could provide the necessary stellar mass measurements \citep[e.g.][]{Barbosa2020OneField, Ferre-Mateu2018OriginsPopulations, Pandya2018ThePhotometry, Greco2018AField, Gu2018LowCluster}.

Finally, most low stellar mass galaxy samples, including ours, are biased towards high density environments. This density bias may further modify the mass-size relation, and may skew the SHMR. Further observations of UDGs in a range of environments and constraints on their GC populations are critical to verify that the discussed trends are real. In future work, we hope to begin reducing the environmental incompleteness of the UDG sample by characterizing the GC populations in a much larger LSBG sample from the \cite{Greco2018IlluminatingSurvey} catalog, encompassing galaxies in a wide range of environments.

\acknowledgements
We'd like to thank Michael Strauss for contributions to the {\it HST} proposal. We'd also like to thank the anonymous referee for  helpful comments. J.P.G. is supported by an NSF Astronomy and Astrophysics Postdoctoral Fellowship under award AST-1801921.

Support for this work was provided by NASA through grant number 15277 from the Space Telescope Science Institute, which is operated by AURA, Inc., under NASA contract NAS 5-26555.

\vspace{5mm}
\facilities{HST(ACS/WFC), Subaru(HSC)}
\software{Astropy \citep{Robitaille2013Astropy:Astronomy, Astropy-Collaboration:2018},
          Drizzlepac \citep{Team2012DrizzlePac:Software, Avila2014DrizzlePacFeatures},
          Matplotlib \citep{Hunter2007Matplotlib:Environment},
          NumPy \citep{Oliphant2006ANumPy, VanDerWalt2011TheComputation},
          Pandas \citep{Team2020Pandas-dev/pandas:Pandas, McKinney2010DataPython},
          SciPy \citep{Virtanen2020SciPyPython},
          statsmodels \citep{Seabold2010Statsmodels:Python},
          PhotUtils \citep{Bradley2019Astropy/photutils:V0.7.2},
          dynesty \citep{Speagle2020Dynesty:Evidences, Skilling2004NestedSampling, Skilling2006NestedComputation}
          }

\bibliography{GCs}{}
\bibliographystyle{aasjournal}

\appendix
\section{GCLF Fitting Procedure}\label{sec:GCLFproc}

\begin{figure}
\gridline{\fig{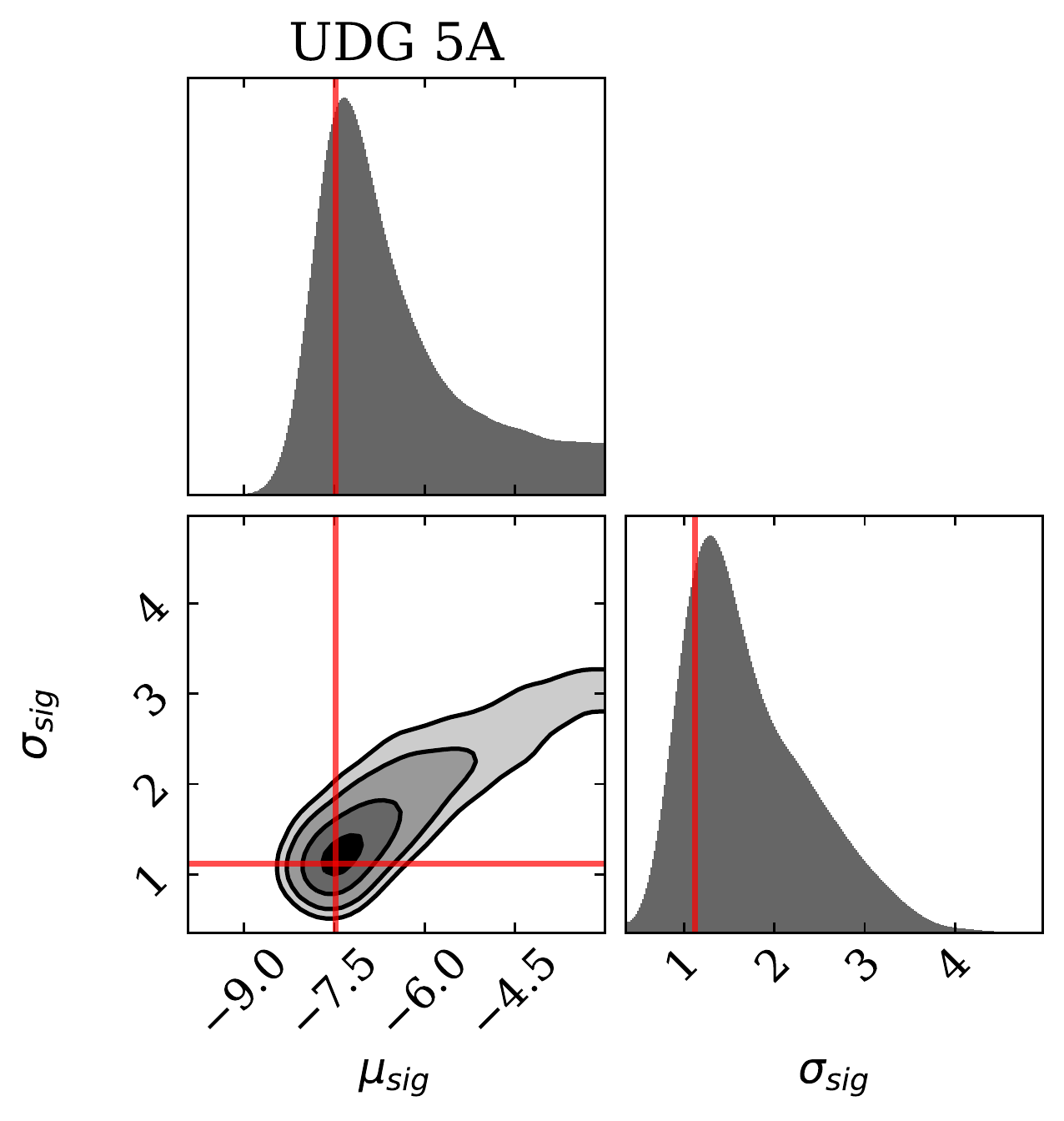}{0.4\textwidth}{},
          \fig{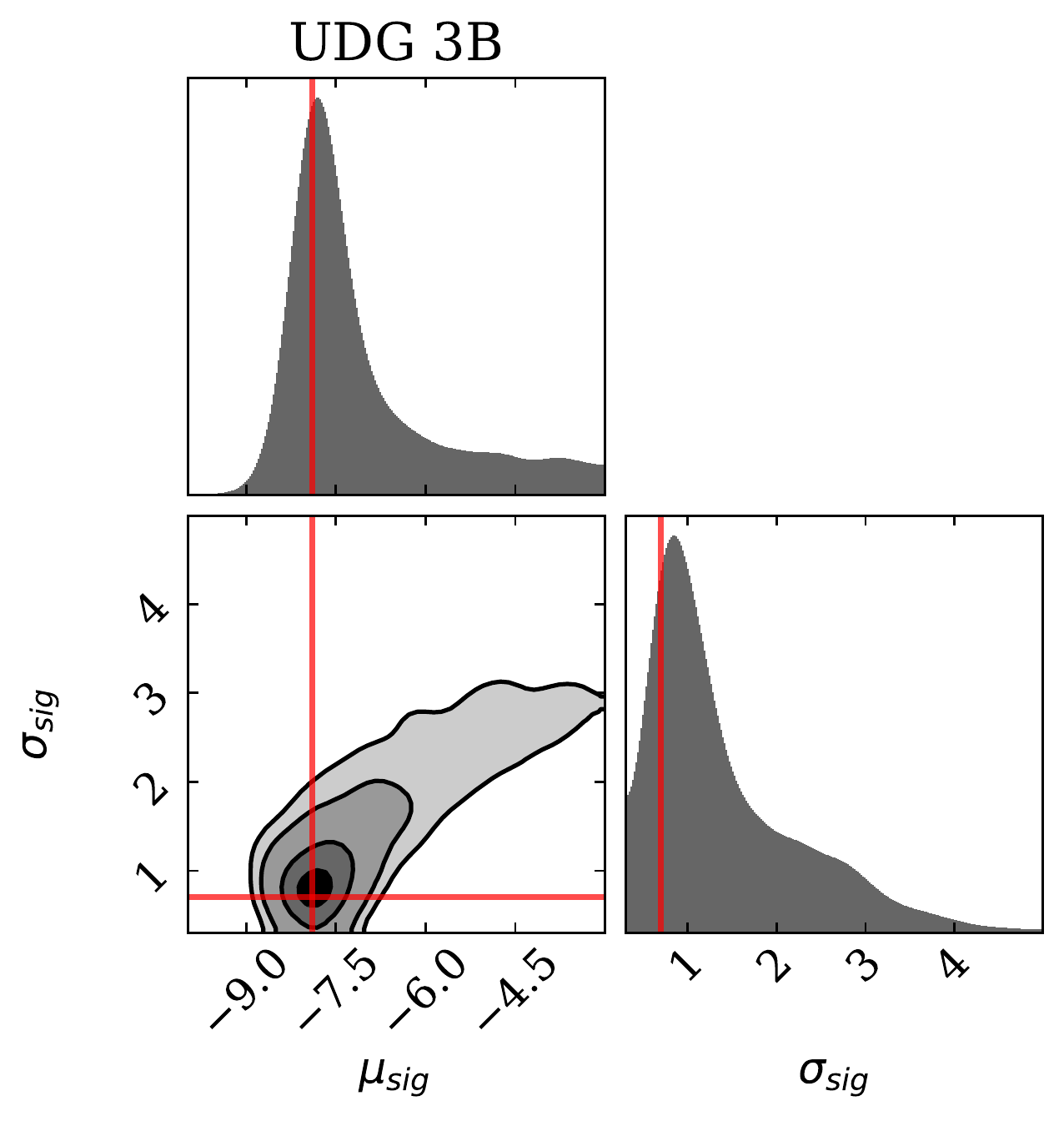}{0.4\textwidth}{}}
\caption{ Corner plots showing the correlation between the GCLF mean ($\mu_{sig}$) and width ($\sigma_{sig}$) for UDGs 5A ({\it left}) and 3B  ({\it right}). Red lines mark the best fit GCLF parameters for each UDG. Although these parameters are correlated, the mean and width are clearly peaked for each UDG. \label{figure:corner_GCLF}}
\end{figure}

In this section, we describe our procedure for fitting the UDG globular cluster luminosity functions. We perform a fit directly to the unbinned data using a maximum likelihood formulation. We choose to use an unbinned fit given the low number of GCs, which would not satisfy $\chi^2$ statistics, and the non-negligible photometric errors which would make the choice of bin size complicated in a binned fit. Moreover, our maximum likelihood framework allows us to easily handle the contaminating background population and our incompleteness in a self-contained way.  For each UDG we fit the individual, unbinned GC magnitudes to the sum of a signal Gaussian and a background distribution. To account for our completeness, we multiply the signal Gaussian by the completeness functions shown in Figure~\ref{figure:completeness}. Thus, for a signal mean and width $\mu_{sig}$ and $\sigma_{sig}$, a completeness $C(v)$ as a function of $V$-band magnitude $v$, and writing a normalized Gaussian as $\mathcal{G}(x\mid \mu, \sigma)$, the probability of observing an actual GC with $V$-band magnitude $v$ is
\begin{gather*}
    P_{sig}(v \mid \mu_{sig}, \sigma_{sig} ) = \frac{\mathcal{G}(v \mid \mu_{sig}, \sigma_{sig} ) C(v)}{\int_{-\infty}^{\infty} \mathcal{G}(v' \mid \mu_{sig}, \sigma_{sig} ) C(v') dv' }.
\end{gather*}
We model the background distribution as a Gaussian, although given the low background level we do not anticipate that variations in this model will affect our results. Given a background mean and width $\mu_{bkg}$ and $\sigma_{bkg}$, the probability of observing a background source of magnitude $v$ is
\begin{gather*}
    P_{bkg}(v \mid \mu_{bkg}, \sigma_{bkg} ) = \mathcal{G}(v \mid \mu_{bkg}, \sigma_{bkg} ).
\end{gather*}

To improve our constraints on the background distribution, we simultaneously fit the galaxy region sources to the combination of the signal and background models and we fit the sources in our background regions to the background distribution. The full model can be summarized by the following log-likelihood, which is a function of the parameters we are most interested in, the signal Gaussian mean and width ($\mu_{sig}$ and $\sigma_{sig}$), but also the background Gaussian mean and width ($\mu_{bkg}$ and $\sigma_{bkg}$), the true total number of sources (including background) in the galaxy region ($\lambda_{tot}$), and the true number of background sources in the galaxy region ($\lambda_{bkg}$). We have written the observed number of sources in the galaxy region as $N_{tot}$, the observed total number of sources in {\it all} of the background regions as $N_{bkg}$, and the number of background regions as $n_{reg}$ (thus, there is an average of $N_{bkg}/n_{reg}$ background sources per background region, and we expect that the true number of background sources in all of the background regions is $\lambda_{bkg} n_{reg}$). We label galaxy region GCs with the index $i$ and background region GCs by the index $j$, and denote the corresponding $V$ band magnitudes $v_i$ and $v_j$.
\begin{gather*} \label{eq:loglike}
    \log\,\mathcal{L}(\mu_{sig}, \sigma_{sig}, \mu_{bkg}, \sigma_{bkg}, \lambda_{tot}, \lambda_{bkg}) = \\
    \Bigg\{ - \lambda_{tot} + \Bigg. \sum_{i=1}^{N_{tot}} \log \Big[ \left(\lambda_{tot} - \frac{\lambda_{bkg}}{n_{reg}}\right) \times P_{sig}(v_i \mid \mu_{sig}, \sigma_{sig} ) \Big. \Bigg.\\
    \Bigg. \Big.+ \left(\frac{\lambda_{bkg}}{n_{reg}}\right) \times P_{bkg}(v_i \mid \mu_{bkg}, \sigma_{bkg} ) \Big] \Bigg\} \\
    + \\
    \Bigg\{ -(n_{reg}\lambda_{bkg})
    + \sum_{j=1}^{N_{bkg}} \log \left[ (n_{reg}\lambda_{bkg}) P_{bkg}(v_j \mid \mu_{bkg}, \sigma_{bkg}) \right] \Bigg\}.
\end{gather*}
A detailed derivation of the log-likelihood for unbinned data in the Poisson regime can be found in Appendix C of \citealp{Drlica-Wagner2020MilkyDR1}. The first part of this equation is the Poisson log-likelihood for the galaxy region model. The second part contains the Poisson log-likelihood describing the model for the sources in all of the background regions. 

We maximize this full log-likelihood to find the best fit $\mu_{sig}$ and $\sigma_{sig}$, although we do float all parameters, and we evaluate the $1\sigma$ errors on these parameters using the \texttt{statsmodels} package \citep[][]{Seabold2010Statsmodels:Python}. The results are described in Section~\ref{sec:GCLF} and shown in Figure~\ref{figure:GCLF}. \changeref{However, there is a known correlation between the GCLF mean and width \citep[][]{Secker1993AEllipticals}, which could limit our ability to constrain those parameters. In Figure~\ref{figure:corner_GCLF}, we show the posterior distributions for the UDG 5A and 3B GCLF means and widths. We evaluated the posterior distribution using the \texttt{dynesty} sampler with a stopping criteria $\Delta \log Z = 0.01$. There is a clear correlation between the mean and width for both GCLFs. However, both are strongly peaked, which suggests that our constraints are meaningful.}


\end{document}